\documentclass[12pt,showpacs,preprintnumbers,superscriptaddress,amsmath,amssymb,nofootinbib]{revtex4}

\usepackage{graphicx}
\usepackage{dcolumn}
\usepackage{bm}
\usepackage{amssymb}
\usepackage{amsmath}
\usepackage{epsfig}    
\usepackage{color}
\usepackage{slashed}

\def\be{\begin{equation}}
\def\ee{\end{equation}}
\newcommand{\bea}{\begin{eqnarray}}
\newcommand{\eea}{\end{eqnarray}}
\newcommand{\nn}{\nonumber}
\newcommand{\eV}{\mbox{eV}}

\newcommand{\GeV}{\mbox{GeV}}
\newcommand{\TeV}{\mbox{TeV}}

\numberwithin{equation}{section}

\begin{document} 
\title{ $T_7$ Flavor Model in Three Loop Seesaw and Higgs Phenomenology }
\preprint{KIAS-P13034}

\author{Yuji Kajiyama}
\email{kajiyama-yuuji@akita-pref.ed.jp}
\affiliation{Akita Highschool, Tegata-Nakadai 1, Akita, 010-0851, Japan}
\author{Hiroshi Okada}
\email{hokada@kias.re.kr}
\affiliation{School of Physics, KIAS, Seoul 130-722, Korea}
\author{Kei Yagyu}
\email{keiyagyu@ncu.edu.tw}
\affiliation{Department of Physics, National Central University, Chungli, Taiwan 32001, ROC}

\begin{abstract}
We propose a new type of radiative seesaw model in which observed neutrino masses are generated through a three-loop level diagram 
in combination with tree-level type-II seesaw mechanism in a  renormalizable theory.
We introduce a Non-abelian flavor symmetry $T_7$ in order to constrain the form of 
Yukawa interactions and Higgs potential.
Although several models based on a Non-abelian flavor symmetry predict the universal coupling constants among  
the standard model like Higgs boson and charged leptons, which is disfavored by the current LHC data, 
our model can avoid such a situation. 
We show a benchmark parameter set that is consistent with the current experimental data, 
and we discuss multi-muon events as a key collider signature to probe our model. 
 
\end{abstract}
\maketitle
\newpage

\section{Introduction}

A new boson has been discovered at Large Hadron Collider (LHC),
whose properties of the production and decay are consistent with those of the Higgs boson in the standard model (SM)~\cite{ATLAS_Higgs,CMS_Higgs}. 
{\it This fact, the observed particle is the SM-like Higgs boson ($h$),  
could seriously affect to models for the charged lepton with flavor symmetries}~\cite{Ishimori:2012zz,Ishimori:2010au,Altarelli:2010gt}, 
{\it since some models could be ruled out.}
As a typical example, we show models based on Non-abelian discrete symmetries such as $A_4$\footnote{The $A_4$ flavor symmetry was initially applied to the lepton sector in Ref. \cite{Babu:2002dz,Ma:2001dn}, in which the structure of the charged lepton does not have the universal coupling. However \cite{Ferreira:2013oga} has those.}, $T_N$ \cite{Luhn:2007sy}, $\Delta(27)$ \cite{Ma:2007wu,Ma:2006ip}, $\Sigma(81)$ \cite{Ishimori:2012gv,Ma:2006ht},
that only have the irreducible representations of singlets (typically introduced as Higgs fields) and triplets (typically as leptons). 
The Lagrangian is then given by
\be
{\cal L}= \sum_{k=1}^3y_k \Phi_k(\bar{L}_ee_R^{} + \omega^{k}\bar{L}_{\mu}\mu_R^{} + \omega^{2k} \bar{L}_{\tau}\tau_R^{})+\text{h.c}.,
\label{univ}
\ee 
where $\Phi_k$, $L_i$ and $(e_R^{},\mu_R^{},\tau_R^{})$ are the isospin Higgs doublet, isospin lepton doublet and isospin lepton singlet fields, respectively, and 
$\omega=\exp(2 \pi i/3)$. 
The mass matrix for the charged leptons $M_\ell$ are then calculated from Eq.~(\ref{univ}) as 
\begin{align}
M_\ell &= \frac{1}{\sqrt{2}}\text{diag}(y_1v_1+y_2v_2+y_3v_3,~y_1v_1+\omega y_2v_2+\omega^2 y_3v_3,~y_1v_1+\omega^2 y_2v_2+\omega y_3v_3)\notag\\
&=\text{diag}(m_e,m_\mu,m_\tau), \label{int2}
\end{align}
where $\langle \Phi_k\rangle =v_k/\sqrt{2}$. 
By solving Eq.~(\ref{int2}), the Yukawa couplings can be derived as 
\begin{align}
&y_1v_1 = \frac{\sqrt{2}}{3}(m_e+m_\mu+m_\tau),~y_2v_2 = \frac{\sqrt{2}}{3}(m_e+\omega^2m_\mu+\omega m_\tau),\notag\\
&y_3v_3 = \frac{\sqrt{2}}{3}(m_e+\omega m_\mu+\omega^2m_\tau). \label{int3}
\end{align}
Although the Lagrangian shown in Eq.~(\ref{univ}) gives the simple structure for the generation of the charged lepton masses as we see in Eqs.~(\ref{int2}) and (\ref{int3}), this results that all the coupling constants among the charged leptons and the CP-even scalar component of $\Phi_k$ are determined by the same Yukawa coupling $y_k$. 
That makes the branching fractions of $h\to \tau^+\tau^-$, $h\to \mu^+\mu^-$ and 
$h\to e^+e^-$ modes to be the same with each other, where $h$ is the mass eigenstate of the CP-even states with the mass of 126 GeV identified to any of $\Phi_k$. 
Such a situation is extremely disfavored by the current results of the Higgs boson search at LHC, namely,  
the event rate for $pp\to h \to \tau^+\tau^-$ is almost the same as that in the SM, while 
the event of $pp\to h \to \mu^+\mu^-$ has not been observed yet\footnote{From the current LHC data, 
$\sigma(pp\rightarrow h \rightarrow \mu^+\mu^-)_{{\rm SM}}\times 10$ has been excluded at 95 $\%$ confidence level~\cite{dimuon}.}.

In the present paper, we clarify the relation between the lepton sector and the Higgs sector in the $T_7$ flavor symmetry~\cite{Ishimori:2012sw, Cao:2011cp, Cao:2010mp,Hagedorn:2008bc}. As for the neutrino sector, two mechanisms of type-II seesaw \cite{typeII} and radiative seesaw with several loops\footnote{
As for the other radiative seesaw models, see Refs.~\cite{zee-babu,Kajiyama:2013zla, Ma:2006km,Sahu, Aoki:2013gzs,Krauss:2002px,Aoki:2008av,Schmidt:2012yg, Bouchand:2012dx, Ma:2012ez, Aoki:2011he, Ahn:2012cg, Farzan:2012sa, Bonnet:2012kz, Kumericki:2012bf,
Kumericki:2012bh, Ma:2012if, Gil:2012ya, Okada:2012np, Hehn:2012kz, Dev:2012sg, Kajiyama:2012xg, Okada:2012sp, Aoki:2010ib, Kanemura:2011vm, Lindner:2011it,
Kanemura:2011mw1,
Kanemura:2011mw2,Kanemura_Sugiyama, Gu:2007ug, Gu:2008zf, Gustafsson, two-triplet, Kanemura:2013qva,Law:2013saa}.} are involved to induce the neutrino observables.
Especially, our scenario requires up to a specific three-loop diagram that gives diagonal components to neutrino mass matrix,
while the off-diagonal components are obtained through the type-II seesaw 
mechanism at tree level.
As for the Higgs sector, we consider the whole Higgs potential and derive all the masses of the Higgs bosons. 
We then analyze the behavior of the SM-like Higgs boson, by fixing a 
benchmark point. 
Our analyses of the Higgs fields could be applied to many models, {\it e.g.} lepton flavor models with Non-abelian discrete symmetries. 
Since the Higgs sector in the present model is similar to that of the Type-X or lepton specific two Higgs doublet model (THDM)~\cite{typeX,typeX2}, 
constraints from hadron collider experiments are rather weak. 
We give a numerical example, in which decay modes of the second lightest 
CP-odd, -even and charged Higgs bosons are mainly muons. This will 
be signals of the present model.  

This paper is organized as follows.
In Section 2, we show particle contents of our model, and discuss 
Higgs boson masses and neutrino masses generated at tree and three- loop level.  
In Section 3, we analyze phenomenology of Higgs bosons. 
We summarize and conclude in Section 4. 
In appendices, some results of detailed calculations for 
the Higgs sector and radiatively induced neutrino masses are given.

\section{Three Loop Radiative Seesaw Model}

In this section, we propose a three-loop radiative seesaw model
which is an extension of the minimal Higgs triplet model motivated from the type-II seesaw mechanism~\cite{typeII}. 
First, we give particle contents and Yukawa interactions. 
After that, masses for the Higgs bosons and neutrinos are discussed. 
\subsection{Model setup}

\begin{table}[thbp]
\centering {\fontsize{10}{12}
\begin{tabular}{c|c|c|c|c|c||c|c|c|c}
\hline\hline ~~Particle~~ & ~~ $L_a$~~ & ~~$e_{aR}$~~& ~~$Q_i$~~& ~~$d_{iR}$~~& ~~$u_{iR}$~~ & ~~ $\Phi_a$~~
  & $\Delta_a$  & $\eta_a$ &$\Phi_q$ \\\hline
$(SU(2)_L,U(1)_Y)$ &  $(\bm{2},-1/2)$ & $(\bm{1},-1)$& $(\bm{2},1/6)$& $(\bm{1},-1/3)$& $(\bm{1},2/3)$  & $(\bm{2},1/2)$  & $(\bm{3},1)$ & $(\bm{2},1/2)$ & $(\bm{2},1/2)$  \\\hline
$T_{7}$   & $\bar3_a$ & $3_a$ &$1_0$&$1_0$&$1_0$ & $\bar 3_a$  & $\bar3_a$ & $3_a$& $1_0$\\\hline
$\mathbb{Z}_2$  & $+$ & $+$ & $+$& $+$& $+$ & $+$  & $+$ & $-$&$+$\\
\hline
$\mathbb{Z}_3$  & $0$ & $0$ &$0$&$-1$&$+1$ & $0$  & $0$ & $0$&$+1$\\
\hline\hline
\end{tabular}%
} \caption{The particle contents, where  $a$ runs 1 to 3. }
\label{tab:b-l}
\end{table}

The particle contents are shown in Tab.~\ref{tab:b-l}. 
We add three $SU(2)_L$ triplet scalar fields $\Delta_a$,
three $SU(2)_L$ doublet scalar fields $\eta_a$, $\Phi_a$ ($a=\tau,e,\mu$), and an  
$SU(2)_L$ doublet scalar field $\Phi_q$,
where $\eta_a$ do not have the vacuum expectation values (VEVs). 
The $\mathbb{Z}_2$ parity is imposed so as to forbid 
terms such as $\bar L\eta e_R$.
The $\mathbb{Z}_3$ parity is introduced in order to avoid couplings among $\Phi_q$ and leptons. 
The renormalizable Lagrangian of Yukawa interactions is given by 
\begin{align}
\mathcal{L}_Y &= y_\ell [\bar{L}_e \Phi_e e_R
+\bar{L}_\mu \Phi_\mu \mu_R+\bar{L}_\tau \Phi_\tau \tau_R]+\text{h.c.}\label{Lyuk}\\
&+ y_d^{ij} \bar{Q}^i\Phi_q d^j_R+ y_u^{ij} \bar{Q}^i\tilde{\Phi}_q u^j_R+\text{h.c.}\label{Lq}\\
&+y_\Delta [(\overline{L_\mu^{ c}}i\tau_2\Delta_\tau+\overline{L_\tau^{c}}i\tau_2\Delta_\mu)L_e
+(\overline{L_\tau^{ c}}i\tau_2\Delta_e+\overline{L_e^{c}}i\tau_2\Delta_\tau)L_\mu\notag\\
&+(\overline{L_e^{c}}i\tau_2\Delta_\mu+\overline{L_\mu^{ c}}i\tau_2\Delta_e)L_\tau]
+\text{h.c.}.  \label{Ldel}
\end{align}
Four doublet Higgs fields and three triplet Higgs fields can be parameterized as 
\begin{align}
\Phi_q &= 
\left[ \begin{array}{c}
\phi_q^+\\
\frac{1}{\sqrt{2}}(h_q+v_q+iz_q)
\end{array}\right],\quad 
\Phi_a = 
\left[ \begin{array}{c}
\phi_a^+\\
\frac{1}{\sqrt{2}}(h_a+v_a+iz_a)
\end{array}\right], \label{vev-1}\\
\Delta_a &=\left[
\begin{array}{cc}
\frac{\Delta_a^+}{\sqrt{2}} & \Delta_a^{++}\\
\Delta_a^0 & -\frac{\Delta_a^+}{\sqrt{2}} 
\end{array}\right],~\text{with}~
\Delta_a^0 =\frac{1}{\sqrt{2}}(\Delta^R_a +v_{\Delta a}+i\Delta^I_a),\quad a=e,\mu,\tau,\label{vev-2}
\end{align}
where $v_a$ and $v_q$ are the VEVs for the doublet fields and $v_{\Delta a}$ are those for the triplet fields,  
which satisfy the sum relation
\begin{align}
\sum_a (v_a^2+2v_{\Delta a}^2)+v_q^2=v^2=(246 \text{ GeV})^2. \label{sum}
\end{align} 
The VEVs of the doublet Higgs fields $\Phi_a$ are related to the charged lepton mass matrix from Eq.~(\ref{Lyuk}) as
\begin{align}
M_\ell = \frac{y_{\ell}}{\sqrt{2}}\text{diag}(v_e,v_\mu,v_\tau)
=\text{diag}(m_e,m_\mu,m_\tau), 
\end{align}
which is already diagonal and has the common Yukawa coupling $y_{\ell}$ 
due to $T_7$ symmetry. 
Mass hierarchy between charged leptons 
should be explained by the hierarchy of VEVs.  
We note that the masses of quarks are generated by $v_q$ in the same way 
as in the SM. 
On the other hand, the triplet VEVs generate neutrino masses. 
Non-zero values for the triplet VEVs cause the deviation in the electroweak rho parameter $\rho$ from unity as
\begin{align}
\rho = \frac{v^2}{v^2+2\sum_a v_{\Delta a}^2}\simeq 1-\frac{2\sum_a v_{\Delta a}^2}{v^2}. \label{rho}
\end{align}
Because the experimental value of $\rho$ is given as $\rho_{\text{exp}}=1.0004^{+0.0003}_{-0.0004}$, 
the triplet VEVs are constrained by $\sum_a v_{\Delta a}^2$ to be about (3.8 GeV)$^2$ at the 95\% confidence level. 
Thus, the sum relation given in Eq.~(\ref{sum}) can be approximately rewritten by 
\begin{align}
v^2\simeq  v_\tau^2+v_q^2.
\end{align} 
The ratio of the above two VEVs can be described as $\tan\beta\equiv v_q/v_\tau$. 

\subsection{Higgs boson masses}

Next, we discuss the Higgs potential, especially for the $\mathbb{Z}_2$ even scalar sector assuming CP conservation.  
The $T_7$ invariant Higgs potential is given by 
\begin{align}
V_{\text{Higgs}}
&=
m_1^2 \Phi^\dagger_a \Phi_a 
+m_2^2 \Phi_q^\dag \Phi_q
+m_3^2{\rm Tr}(\Delta^\dag_a\Delta_a)
\notag\\&
+\mu(\Phi_\mu^T i\tau_2\Delta_\tau^\dag \Phi_\mu+\Phi_\tau^Ti\tau_2 \Delta_e^\dag \Phi_\tau+\Phi_e^T i\tau_2\Delta_\mu^\dag \Phi_e+\text{h.c.})
\notag\\&
+\mu_\eta (\eta^T_1i\tau_2 \Delta_2^\dag \eta_3+\eta^T_2i\tau_2 \Delta_3^\dag \eta_1
+\eta^T_3 i\tau_2\Delta_1^\dag \eta_2+\text{h.c.})
\notag\\
&+
\lambda_{\ell1} (\Phi^\dagger_a \Phi_a)^{2}
+\lambda_q(\Phi_q^\dag \Phi_q)^2
+\lambda_{q\ell1} (\Phi_a^\dag \Phi_a)(\Phi_q^\dag \Phi_q)\notag\\
&
+\lambda_\pm (|\Phi^\dagger_\tau \Phi_e|^2
+|\Phi^\dagger_e \Phi_\mu|^2+|\Phi^\dagger_\mu \Phi_\tau|^2 )
+\lambda_{q\ell2} |\Phi_a^\dag \Phi_q|^2\notag\\
&+\lambda_{\ell2}  (\Phi^\dagger_\tau \Phi_\tau+\omega^2\Phi^\dagger_e \Phi_e
+\omega\Phi^\dagger_\mu \Phi_\mu)(\Phi^\dagger_\tau \Phi_\tau+\omega\Phi^\dagger_e \Phi_e
+\omega^2\Phi^\dagger_\mu \Phi_\mu), \label{pot1}
\end{align} 
where terms with the index $a$ should be summed over $a=\tau,e,\mu$. 
In addition to the above terms, 
we introduce the following soft terms which break the $T_7$ symmetry into $\mathbb{Z}_3$ symmetry;  
\begin{align}
V_{\text{soft}}&=
\tilde m_{1a}^2 \Phi_a^\dag \Phi_a 
+\tilde m_{3a}^2 {\rm Tr}(\Delta^{\dag}_a \Delta_a)
+\tilde m_{ \Phi}^2 ( \Phi_e^\dag \Phi_\mu
+\Phi_\mu^\dag \Phi_\tau+\Phi_\tau^\dag \Phi_e+\text{h.c.}  )\notag\\&
+\tilde{\mu}(\Phi^T_q i\tau_2\Delta_a^{\dag}\Phi_q+\text{h.c.} ). 
\label{pot2}
\end{align}
The potential given in Eqs.~(\ref{pot1}) and (\ref{pot2}) is not the most general form, where 
we only write down terms which are relevant to the following phenomenological studies\footnote{
Most of terms which are not displayed in Eqs.~(\ref{pot1}) and (\ref{pot2}) are not important from the following reasons. 
First, they can contribute to masses for scalar bosons with the magnitude of $\mathcal{O}(v_\Delta^2)$. Such a contribution 
can be negligible, because of $v_\Delta\ll v$.  Second, they give masses for $\mathbb{Z}_2$ odd scalar bosons, which are not related to the 
following discussions. }. 
The complete expressions for the Higgs potential are given in Appendix~A. 
We note that 
the $\tilde{\mu }$ term is necessary to break an accidental $U(1)$ symmetry of 
the phase rotation $\Phi_q \to e^{i \theta}\Phi_q$, otherwise there exists an additional 
massless Nambu-Goldstone (NG) boson that cannot be absorbed by a gauge field.
It implies that the mass scale of $\tilde{\mu }$ must be ${\cal O}(100)$ GeV at most, since 
$\tilde{\mu }$ plays an important role in obtaining mass of the lightest CP-odd Higgs boson 
 in Eq.~(\ref{cp-odd}). The LHC experiment tells us that its mass must be more than 100 GeV 
as we will discuss in the section~\ref{higgses}.
%

From the tadpole condition, we obtain
\begin{align}
\tilde{m}_{1e}^2+m_1^2&=
-\frac{v_\mu+v_\tau }{v_e}\tilde m_{ \Phi}^2-\frac 12\left[\lambda_{q\ell}v_q^2
+2 \lambda_\ell v_e^2
+ (2\lambda_{\ell 1}-\lambda_{\ell 2}+\lambda_\pm)
\left( v_\mu^2+v_\tau^2\right)\right]
+{\sqrt{2}}\mu  v_{\Delta \mu},\notag\\
\tilde{m}_{1\mu}^2+m_1^2&=
-\frac{v_\tau +v_e}{v_\mu}\tilde m_{ \Phi}^2-\frac 12\left[\lambda_{q\ell}v_q^2
+2 \lambda_\ell v_\mu^2
+(2\lambda_{\ell 1}-\lambda_{\ell 2}+\lambda_\pm)
\left( v_\tau ^2+v_e^2\right)\right]
+{\sqrt{2}}\mu  v_{\Delta \tau },\notag\\
\tilde{m}_{1\tau}^2+m_1^2&=
-\frac{v_e+v_\mu}{v_\tau }\tilde m_{ \Phi}^2 -\frac 12\left[ \lambda_{q\ell}v_q^2
+2 \lambda_\ell v_\tau ^2
+(2\lambda_{\ell 1}-\lambda_{\ell 2}+\lambda_\pm)
\left( v_e^2+v_\mu^2\right)\right]
+{\sqrt{2}}\mu  v_{\Delta e},\notag\\
m_2^2&=-\lambda_qv_q^2
-\frac 12\lambda_{q\ell}\left( v_\tau ^2+v_e^2+v_\mu^2\right)
+\sqrt{2}\tilde{\mu}\left( v_{\Delta \tau }+v_{\Delta e}+v_{\Delta \mu}\right),\notag\\
\tilde{m}_{3e}^2+m_3^2&=
\frac{\sqrt{2}(\mu  v_\tau^2+\tilde{\mu} v_q^2)}{2v_{\Delta e}},\notag\\
\tilde{m}_{3\mu}^2+m_3^2&=
\frac{\sqrt{2}(\mu v_e^2+\tilde{\mu} v_q^2)}{2v_{\Delta \mu}},\notag\\
\tilde{m}_{3\tau}^2+m_3^2&=
\frac{\sqrt{2}(\mu  v_\mu^2+\tilde{\mu} v_q^2)}{2v_{\Delta \tau}}, 
\label{tadpole}
\end{align}
where $\lambda_\ell\equiv \lambda_{\ell1}+\lambda_{\ell2}$ and $\lambda_{q\ell}\equiv \lambda_{q\ell1}+\lambda_{q\ell2}$.

There are seven (six) physical CP-even (CP-odd) scalar bosons, 
six pairs of singly-charged scalar bosons and 
three pairs of the doubly-charged scalar bosons in addition to the neutral $G^0$ and the charged $G^\pm$ NG bosons 
which are obtained after the diagonalization and absorbed by the longitudinal component of 
the $Z$ and $W$ bosons, respectively. 
The mass matrix for the CP-even Higgs bosons in the basis of $(h_q,h_\tau,h_\mu,h_e,\Delta_\tau^R,\Delta_\mu^R,\Delta_e^R)$ is given by 
\begin{align}
M_{R}^2&=
\left[
\begin{array}{cc}
M_{\Phi_R}^2 & M_{\Phi_R\Delta_R}^2 \\
(M_{\Phi_R\Delta_R}^2)^T & M_{\Delta_R}^2
\end{array}
\right], 
\end{align}
where $M_{\Phi_R}^2$, $M_{\Delta_R}^2$ and $M_{\Phi_R\Delta_R}^2$ are  
the sub-matrices of $M_{R}^2$, which are $4\times 4$, $3\times 3$ and $4\times 3$ forms, respectively. 
They can be expressed as
\begin{align}
M_{\Phi_R}^2&\simeq v^2\left[
\begin{array}{cccc}
2s_\beta^2\lambda_q & s_\beta c_\beta \lambda_{ q\ell} & s_\beta c_\beta\epsilon_\mu \lambda_{q\ell} & s_\beta c_\beta\epsilon_e \lambda_{q\ell} \\
s_\beta c_\beta \lambda_{ q\ell}& -\frac{\tilde{m}_{ \Phi}^2}{v^2}\epsilon_\mu+2c_\beta^2 \lambda_{\ell} & \frac{\tilde{m}_{ \Phi}^2}{v^2}  & \frac{\tilde{m}_{ \Phi}^2}{v^2} \\
s_\beta c_\beta\epsilon_\mu \lambda_{q\ell}&\frac{\tilde{m}_{ \Phi}^2}{v^2}&  -\frac{\tilde{m}_{ \Phi}^2}{v^2}\frac{1}{\epsilon_\mu} & \frac{\tilde{m}_{ \Phi}^2}{v^2}\\
s_\beta c_\beta\epsilon_e \lambda_{q\ell}&\frac{\tilde{m}_{ \Phi}^2}{v^2}&\frac{\tilde{m}_{ \Phi}^2}{v^2}& -\frac{\tilde{m}_{ \Phi}^2}{v^2}\frac{1}{\epsilon_e}
\end{array}\right],\notag\\
M_{\Delta_R}^2 &\simeq
v^2\left[
\begin{array}{ccc}
\frac{\tilde{\mu}s_\beta^2+\mu c_\beta^2\epsilon_\mu^2}{\sqrt{2}v_{\Delta \tau}}
&0&0\\
0&\frac{\tilde{\mu}s_\beta^2+\mu c_\beta^2\epsilon_e^2}{\sqrt{2}v_{\Delta \mu}}&0 \\
0&0&\frac{\tilde{\mu}s_\beta^2+\mu c_\beta^2}{\sqrt{2}v_{\Delta e}}\\
\end{array}\right],~~
M_{\Phi_R\Delta_R}^2
\simeq -v^2\left[
\begin{array}{ccc}
\frac{\sqrt{2}\tilde{\mu}s_\beta}{v}& \frac{\sqrt{2}\tilde{\mu}s_\beta}{v} & \frac{\sqrt{2}\tilde{\mu}s_\beta}{v} \\
0&0&\frac{\sqrt{2}\mu c_\beta}{v} \\
\frac{\sqrt{2}\mu c_\beta\epsilon_\mu}{v} &0&0 \\
0&\frac{\sqrt{2}\mu c_\beta\epsilon_e}{v} &0
\end{array}\right],  \label{MRsq}
\end{align}
where $\epsilon_{\mu,e}=m_{\mu,e}/m_\tau$.
Similarly, each of the mass matrix for the CP-odd scalar bosons 
and that for the singly-charged scalar bosons in the basis of 
$(z_q,z_\tau,z_\mu,z_e,\Delta_\tau^I,\Delta_\mu^I,\Delta_e^I)$
and $(\phi_q^\pm,\phi_\tau^\pm,\phi_\mu^\pm,\phi_e^\pm,\Delta_\tau^\pm,\Delta_\mu^\pm,\Delta_e^\pm)$
can be expressed by 
\begin{align}
M_I^2&=
\left[
\begin{array}{cc}
M_{\Phi_I}^2 & M_{\Phi_I\Delta_I}^2 \\
(M_{\Phi_I\Delta_I}^2)^T & M_{\Delta_I}^2
\end{array}
\right], \quad 
M_{\pm}^2=
\left[
\begin{array}{cc}
M_{\Phi^+}^2 & M_{\Phi^+\Delta^+}^2 \\
(M_{\Phi^+\Delta^+}^2)^T & M_{\Delta^+}^2
\end{array}
\right], 
\end{align}
where $M_{\Phi_I}^2$ ($M_{\Phi_+}^2$), $M_{\Delta_I}^2$ ($M_{\Delta_+}^2$) 
and $M_{\Phi_I\Delta_I}^2$ ($M_{\Phi_+\Delta_+}^2$) are 
the sub-matrix for $M_I^2$ ($M_+^2$) which are $4\times 4$, $3\times 3$ and $4\times 3$, respectively. 
Each sub-matrix can be obtained as 
\begin{align}
M_{\Phi_I}^2 &\simeq 
\left[
\begin{array}{cccc}
2\sqrt{2}\tilde{\mu}\sum_a^{e,\mu,\tau} v_{\Delta a}&0&0&0\\
0&-\tilde{m}_{ \Phi}^2\epsilon_\mu +2\sqrt{2}\mu v_{\Delta e}&\tilde{m}_{ \Phi}^2&\tilde{m}_{ \Phi}^2\\
0&\tilde{m}_{ \Phi}^2&-\frac{\tilde{m}_{ \Phi}^2}{\epsilon_\mu} +2\sqrt{2}\mu v_{\Delta \tau}&\tilde{m}_{ \Phi}^2\\
0&\tilde{m}_{ \Phi}^2&\tilde{m}_{ \Phi}^2&-\frac{\tilde{m}_{ \Phi}^2}{\epsilon_e}+2\sqrt{2}\mu v_{\Delta \mu}
\end{array}\right],   \notag\\
M_{\Delta_I}^2 &= M_{\Delta_R}^2,\quad M_{\Phi_I\Delta_I}^2 = M_{\Phi_R\Delta_R}^2, \label{cp-odd}
\end{align}
and 
\begin{align}
M_{\Phi^+}^2&\simeq \frac{1}{2}\lambda_\pm v^2\left[
\begin{array}{cccc}
-c_\beta^2(1+\epsilon_\mu^2+\epsilon_e^2)& \frac{1}{2}s_{2\beta} & \frac{1}{2}s_{2\beta}\epsilon_\mu  & \frac{1}{2}s_{2\beta} \epsilon_e  \\
\frac{1}{2}s_{2\beta}& -s_\beta^2 & 0  & 
0\\
\frac{1}{2}s_{2\beta}\epsilon_\mu&0&  -s_\beta^2 & 0\\
\frac{1}{2}s_{2\beta} \epsilon_e&0&0& -s_\beta^2
\end{array}\right]+M_{\Phi_I}^2(\mu\to\mu/2,\tilde{\mu}\to\tilde{\mu}/2),\notag\\
M_{\Delta^+}^2&=M_{\Delta_R}^2,\quad
M_{\Phi^+\Delta^+}^2
=\frac{1}{\sqrt{2}}M_{\Phi_R\Delta_R}^2.
\end{align}
One can find that the eigenvectors belonging to the NG modes $G^0$ and $G^\pm$ 
are given as 
\begin{align}
\vec{v}_{G^0} &= (v^2+2\sum_a v_{\Delta a}^2)^{-1/2}(v_q,v_\tau,v_\mu,v_e,2v_{\Delta\tau},2v_{\Delta\mu},2v_{\Delta e}),\\
\vec{v}_{G^+} &= v^{-1}(v_q,v_\tau,v_\mu,v_e,\sqrt{2}v_{\Delta\tau},\sqrt{2}v_{\Delta\mu},\sqrt{2}v_{\Delta e}).
\end{align}
We can construct the unitary matrices which make $M_{I}^2$ and $M_{+}^2$ 
to be the block diagonal forms by using $\vec{v}_{G^0}$, $\vec{v}_{G^+}$ and those orthogonal vectors, 
in which the NG modes are decoupled from the physical scalar states.
We do not show explicitly these unitary matrices, because their analytic formulae are too complicated to show in the paper. 

Under $v_\Delta \ll v$ which is required by the rho parameter data,  
the diagonal elements of $M_{\Delta_R}^2$ become much larger than 
elements in $M_{\Phi_R\Delta_R}^2$. 
In that case,  
the diagonalization matrices for $M_{R}^2$, $M_{I}^2$ and $M_+^2$ 
are given as a block diagonal form like $4\times 4$ and $3\times 3$, so that 
we can separately consider the mass eigenstates which are mainly composed of doublets from those of triplets. 
We then define the mass eigenstates for the CP-even, CP-odd and singly-charged scalar bosons as follows
\begin{align}
\left(
\begin{array}{c}
h_q\\
h_\tau\\
h_\mu \\
h_e 
\end{array}\right)=U_{R}
\left(
\begin{array}{c}
h\\
H_\tau\\
H_\mu \\
H_e 
\end{array}\right),~
\left(
\begin{array}{c}
z_q\\
z_\tau\\
z_\mu \\
z_e 
\end{array}\right)=U_{I} 
\left(
\begin{array}{c}
G^0\\
A_\tau\\
A_\mu \\
A_e 
\end{array}\right),~
\left(
\begin{array}{c}
\phi_q^+\\
\phi_\tau^+ \\
\phi_\mu^+ \\
\phi_e^+ 
\end{array}\right)=U_{+} 
\left(
\begin{array}{c}
G^+\\
H_\tau^+\\
H_\mu^+ \\
H_e^+ 
\end{array}\right). 
\end{align}
The masses for the Higgs bosons can be calculated as 
\begin{align}
&U_R^TM_{\Phi_R}^2U_R=\text{diag}(m_{h}^2,m_{H_\tau}^2,m_{H_\mu}^2,m_{H_e}^2),\\
&U_I^TM_{\Phi_I}^2U_I=\text{diag}(0,m_{A_\tau}^2,m_{A_\mu}^2,m_{A_e}^2),\\
&U_+^TM_{\Phi^+}^2U_+=\text{diag}(0,m_{H_\tau^+}^2,m_{H_\mu^+}^2,m_{H_e^+}^2). 
\end{align}
Approximately, the masses for the Higgs bosons can be expressed in the case of $v_{\Delta\tau}=v_{\Delta\mu}=v_{\Delta e}=v_\Delta$ as
\begin{align}
&m_h^2\simeq(\bar{M}_{\Phi_R}^2)_{11}c_\alpha^2+(\bar{M}_{\Phi_R}^2)_{22}s_\alpha^2+2(\bar{M}_{\Phi_R}^2)_{12}c_\alpha s_\alpha,\\
&m_{H_\tau}^2\simeq(\bar{M}_{\Phi_R}^2)_{11}s_\alpha^2+(\bar{M}_{\Phi_R}^2)_{22}c_\alpha^2-2(\bar{M}_{\Phi_R}^2)_{12}c_\alpha s_\alpha,\\
&m_{A_\tau}^2 \simeq 
2\sqrt{2}\mu v_\Delta \left[1-\frac{\mu c_\beta^2}{\tilde{\mu}s_\beta^2+\mu c_\beta^2}\right],\\
&m_{H_\tau^+}^2 \simeq
\sqrt{2}\mu v_\Delta-\frac{\lambda_\pm}{2}v^2s_\beta^2
 -\frac{\sqrt{2}v_\Delta }{\tilde{\mu}s_\beta^2+\mu c_\beta^2}\Big[\frac{\lambda_\pm^2}{8}v^2s_\beta^4+\mu ^2c_\beta^2\Big],\\
&m_{H_{\mu,e}}^2\simeq m_{A_{\mu,e}}^2\simeq m_{H_{\mu,e}^+}^2\simeq -\frac{\tilde{m}^2_{ \Phi}}{\epsilon_{\mu,e}},
\end{align}
with
\begin{align}
&(\bar{M}_{\Phi_R}^2)_{11}=
2v^2s_\beta^2\Big[\lambda_q-\sqrt{2}v_\Delta(s_\beta^2\lambda_q^2+\frac{\tilde{\mu}^2}{v^2})\Big(\frac{1}{\tilde{\mu}s_\beta^2+\mu c_\beta^2}+\frac{2}{\tilde{\mu}s_\beta^2}\Big)\Big],\\
&(\bar{M}_{\Phi_R}^2)_{22}=
-\tilde{m}_{ \Phi}^2\epsilon_\mu+2v^2c_\beta^2\left[\lambda_\ell-\sqrt{2}v_\Delta\frac{c_\beta^2\lambda_\ell^2+\mu ^2/v^2}{\tilde{\mu}s_\beta^2+\mu c_\beta^2}\right],\\
&(\bar{M}_{\Phi_R}^2)_{12}=v^2c_\beta s_\beta\Big[\lambda_{q\ell}-\sqrt{2}v_\Delta\Big(\frac{1}{2}s_\beta c_\beta\lambda_{q\ell}^2+\frac{2\tilde{\mu}^2t_\beta}{v^2}\Big)\Big(\frac{1}{\tilde{\mu}s_\beta^2+\mu c_\beta^2}+\frac{2}{\tilde{\mu}s_\beta^2}\Big)\Big],\\
&\tan2\alpha = -\frac{2(M_{\Phi_R}^2)_{12}}{(M_{\Phi_R}^2)_{11}-(M_{\Phi_R}^2)_{22}}. 
\end{align} 
In addition, first $2\times 2$ part of the unitary matrices are written as
\begin{align}
&U_{R}^{11} \simeq c_\alpha,\quad U_{R}^{12}\simeq s_\alpha,\quad U_{R}^{21}\simeq-s_\alpha,\quad U_{R}^{22} \simeq c_\alpha\\
&U_{I,+}^{11}\simeq s_\beta,\quad U_{I,+}^{12}\simeq c_\beta, \quad U_{I,+}^{21}\simeq c_\beta,\quad 
U_{I,+}^{22}\simeq -s_\beta. 
\end{align}
We note that $m_{A_\tau}^2$ is 
dominantly proportional to $\mu  v_\Delta$, so that $\mu $ should be taken to be as 
large as $\mathcal{O}(1)$ TeV and $v_\Delta$ to be of $\mathcal{O}(1)$ 
GeV to compensate the tiny triplet VEV.   
Moreover, if one assumes that the mass of $\Delta$ is of ${\cal O}(1)$ TeV, $\mu$ 
should be at most of ${\cal O}(1)$ TeV, because there is strong correlation between $v_{\Delta}$ and $\mu $, $ v_{\Delta} \sim \mu  v^2/m_{\Delta}^2$, as is often the case with the usual type-II seesaw.  

The mass matrix for the doubly-charged scalar states is expressed by the $3\times 3$ form, because they purely come 
from the triplet Higgs fields. 
In the basis of $(\Delta_\tau^{\pm\pm},\Delta_\mu^{\pm\pm},\Delta_e^{\pm\pm})$, the mass matrix is the same as
$M_{\Delta_R}^2$ given in Eq.~(\ref{MRsq}). 
Their squared mass eigenvalues are typically determined by $(\mu+\tilde{\mu})v^2/v_\Delta$, so that these 
are ($\mathcal{O}(1)$ TeV)$^2$ as long as we take $v_\Delta=\mathcal{O}(1)$ GeV and $\mu,\tilde{\mu}=\mathcal{O}(1)$ TeV. 

In the following discussion, we treat $h$ as the SM-like Higgs boson which should be identified to be a
new boson discovered at LHC with the mass of $126$ GeV. 


\subsection{Neutrino mass matrix}
\begin{figure}[t]
\begin{center}
 \includegraphics[width = 100mm]{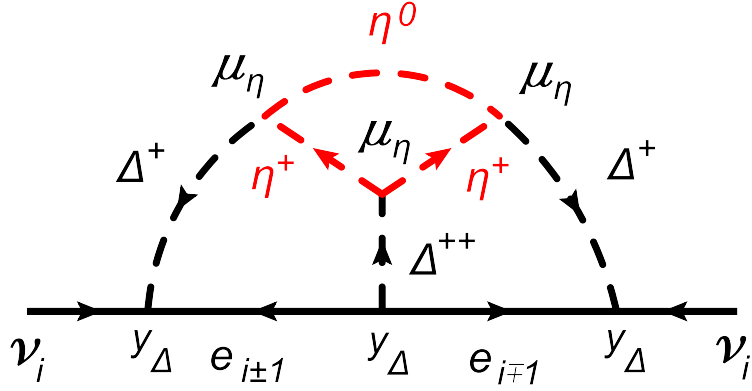}
   \caption{Neutrino mass generation via the three-loop level diagram. 
The particles indicated by a red font have the opposite $\mathbb{Z}_2$ charge to those by a black font. }
   \label{three-neut}
\end{center}
\end{figure}
In this subsection, we discuss neutrino masses which are induced at three-loop level as well as at tree level. 
First, the tree-level mass matrix for neutrinos $M_{\nu}^{(0)}$ is given through Eq.(\ref{Ldel}) as
\be
M_{\nu}^{(0)}=\frac{y_{\Delta}}{\sqrt{2}}
\left(\begin{array}{ccc}
0&v_{\Delta \tau}&v_{\Delta \mu}\\
v_{\Delta \tau}&0&v_{\Delta e}\\
v_{\Delta \mu}&v_{\Delta e}&0\\
\end{array}\right).
\label{mnu0}
\ee
When $v_{\Delta \tau}=v_{\Delta \mu}$, 
the Pontecorvo-Maki-Nakagawa-Sakata matrix at the tree level $U^{(0)}$ is given by 
\be
U^{(0)}=\left( \begin{array}{ccc} 
-(a_e+\sqrt{a_{e\tau}})/\sqrt{+}&-(a_e-\sqrt{a_{e\tau}})/\sqrt{-}&0\\
2a_\tau/\sqrt{+}&2a_\tau/\sqrt{-}&-1/\sqrt{2}\\
2a_\tau/\sqrt{+}&2a_\tau/\sqrt{-}&1/\sqrt{2}\\
\end{array}\right), 
\ee
and it diagonalizes $M_{\nu}^{(0)}$ as 
\be
U^{(0)T}M_{\nu}^{(0)}U^{(0)}=
{\rm diag}\left( (a_e-\sqrt{a_{e\tau}})/2,~(a_e+\sqrt{a_{e\tau}})/2,~-a_e\right), 
\ee
where 
\be
a_\ell=\frac{y_{\Delta}}{\sqrt{2}}v_{\Delta \ell},~
a_{e\tau}\equiv a_e^2+8 a_\tau^2,~
\sqrt{\pm}=\sqrt{8 a_\tau^2+(a_e \pm \sqrt{a_{e\tau}})^2}.
\ee
$U_{e3}^{(0)}=0$ and the maximal mixing are originated from the condition 
$v_{\Delta \tau}=v_{\Delta \mu}$. However, this condition can 
be relaxed to obtain non-zero $U_{e3}$ and observed value of $\theta_{12(23)}$ 
at $3\sigma$ range. 
As mentioned before, we take $v_{\Delta}= {\cal O}(1)~\GeV$ to obtain the phenomenologically 
enough large mass for $A_\tau$, so that 
$y_{\Delta}= {\cal O}(10^{-10})$ is required to reproduce $m_{\nu}= {\cal O}(0.1)~\eV$. 
Since there exist only three parameters in Eq.~(\ref{mnu0}), 
one cannot derive all observables in neutrino sector. 
Therefore, one has to take loop-level mass matrices into account.

Next, we discuss loop level neutrino masses. The tree level mass matrix does 
not have non-zero diagonal entries and neutrino observables cannot be 
derived at tree level. Therefore we focus on diagonal elements induced by 
loop diagrams. To achieve it, we summarize how to generate Majorana neutrino mass 
matrix by Yukawa interactions given in Eqs. (\ref{Lyuk}) and (\ref{Ldel})
as listed in the following:
\begin{description}
\item{(i)} Each $\Delta$ attached on the fermion line gives one $y_{\Delta}={\cal O}(10^{-10})$. 
\item{(ii)} Odd number of $\Delta$s should be attached on the fermion line for lepton number violation.
\item{(iii)} At least, three $\Delta$s should be attached on the fermion line 
in order to generate diagonal elements of neutrino mass matrix, because 
$\Delta$ changes the lepton flavor, while $\Phi$ does not. 
If only one $\Delta$ is attached, it always gives off-diagonal elements of 
neutrino mass matrix.  
\item{(iv)} On fermion line between $\Phi$ and $\Delta$, chirality suppression occurs. 
\end{description}
Along these lines of thought, one finds that the sizable contributions to 
diagonal elements are 
given at three-loop level shown in Fig. \ref{three-neut},  
and its magnitude can be 
estimated as\footnote{The exact form is found in Appendix \ref{three-neutrino-mass}.} 
\be
M_{\nu}^{(3)}\sim \frac{1}{(16 \pi^2)^3}y_{\Delta}^3 \mu_\eta^3 
\frac{m_{\eta}^2}{m_{\Delta}^4}
= {\cal O}(0.1)~\eV, 
\ee
where $\mu_\eta$ is the dimensionful coupling associated with the $T_7$ invariant term defined in Eq.~(\ref{pot1}), 
and $m_{\eta}$ and $m_{\Delta}$ are the typical $\eta$ and $\Delta$ masses; $m_{\eta}
={\cal O}(10^8)~\GeV$
and $m_{\Delta}={\cal O}(1)~\TeV$, respectively. 
Notice here that the mass parameter $\mu_\eta$, which is taken to be ${\cal O}(10^8)~
\GeV$, contributes only to the 
$\eta$ masses, since  $\eta$ do not mix with $\Phi$ and $\Delta$ due to the inert property.
Hence one can take arbitrarily large value for $\mu_\eta$ with no effects to masses of 
$\Phi$ and $\Delta$
\footnote{This is, in a sense, a fine-tuning that one has to tune the bare mass of $\Delta$ to cancel the large loop contribution from $\eta$ in order to obtain the small physical mass of $\Delta$. 
This is similar to the renormalization of the Higgs boson mass in the SM if the cutoff 
scale is assumed to be a large scale such as the grand unification scale or 
Planck scale.
We expect that such a fine-tuning problem may be able to be solved by extending our model to the supersymmetric theory. We would like to thank the referee to draw our attention to such kind of matter.}.
As a result, we can reproduce observed neutrino masses $\sim{\cal O}$(0.1) eV  and the mixing data because of many parameters, as can be seen in Appendix 
\ref{three-neutrino-mass}. As for the other contributions up to three-loops, see Appendix \ref{minor-cont}.

\section{Higgs Phenomenologies}\label{higgses}

In this section, we discuss the collider phenomenology of the Higgs bosons. 
This model can be effectively regarded as the so-called Type-X or lepton specific THDM~\cite{typeX,typeX2}, in which 
one of the two Higgs doublets couples to quarks and the other one couples to leptons.  
To see this, we write down the interaction terms in the Yukawa Lagrangian given in Eqs.~(\ref{Lyuk}) and (\ref{Lq})
in terms of the mass eigenstates of the Higgs bosons as follows
\begin{align}
&\mathcal{L}_Y=\notag\\
&\frac{m_\tau}{vc_\beta}\sum_{\ell=e,\mu,\tau}\Big[\bar{\ell}\ell (hU_{R}^{\ell 1}+
H_\tau U_{R}^{\ell 2}+H_\mu U_{R}^{\ell 3}+H_e U_{R}^{\ell 4})
+i\bar{\ell}\gamma_5\ell (
A_\tau U_{I}^{\ell 2}+A_\mu U_{I}^{\ell 3}+A_e U_{I}^{\ell 4})\Big]\notag\\
&+\frac{\sqrt{2}m_\tau}{vc_\beta}[\bar{\nu}_\ell P_R \ell (H_\tau^+ U_{+}^{\ell 2}+H_\mu^+ U_{+}^{\ell 3}+H_e^+ U_{+}^{\ell 4})+\text{h.c.}]\notag\\
&+\frac{m_q}{vs_\beta}\Big[\bar{q}q (hU_{R}^{11}+
H_\tau U_{R}^{12}+H_\mu U_{R}^{13}+H_e U_{R}^{14})
+\text{Sign}(q)i\bar{q}\gamma_5 q(
A_\tau U_{I}^{12}+A_\mu U_{I}^{13}+A_e U_{I}^{14})\Big]\notag\\
&+\frac{\sqrt{2}}{vs_\beta}\Big[
\bar{u} (m_d P_R-m_uP_L) d(H_\tau^+ U_{+}^{12}+H_\mu^+ U_{+}^{13}+H_e^+ U_{+}^{14})+\text{h.c.}\Big], 
\end{align}
where Sign($q$)=1~($-1$) for $q=d$ ($q=u$), and the projection operators are $P_{L}
=(1-\gamma_5)/2$ and $P_{R}=(1+\gamma_5)/2$. 
From the above expression, the ratio of the Yukawa coupling in our model to that in the SM, $c_{hff}$, and the similar rate of the 
gauge coupling $c_{hVV}$ for $h$ can be calculated as
\begin{align}
&c_{hq\bar{q}}=\frac{U_{R}^{11}}{s_\beta}\simeq \frac{c_\alpha}{s_\beta},~
c_{h\tau\tau}=\frac{U_{R}^{21}}{c_\beta}\simeq -\frac{s_\alpha}{c_\beta},~
c_{h\mu\mu}=\frac{U_{R}^{31}}{c_\beta \epsilon_\mu},~
c_{hee}=\frac{U_{R}^{41}}{c_\beta \epsilon_e},\notag\\
&c_{hVV}=s_\beta U_{R}^{11}+c_\beta U_{R}^{21}\simeq \sin(\beta-\alpha). 
\end{align}
The SM-like limit, in which the coupling constants of $h$ are the same as those in the SM Higgs boson, 
can be obtained by taking the following limit
\begin{align}
U_{R}^{11} \to s_\beta,\quad U_{R}^{21} \to c_\beta.\label{SMlike}
\end{align}
By using $\alpha$, this limit can also be approximately expressed as $\sin(\beta-\alpha)\to 1$
which is the same form known in the THDMs with a softly broken discrete $\mathbb{Z}_2$ symmetry. 
In the SM-like limit, 
factors of the vertices among the lightest extra Higgs bosons ($H_\tau$, $A_\tau$ and $H_\tau^\pm$) 
and fermions are calculated by 
\begin{align}
&(H_\tau \tau\tau): \frac{U_R^{22}}{c_\beta}\simeq \tan\beta,~~
(H_\tau u\bar{u}):-\frac{U_R^{12}}{c_\beta}\simeq \cot\beta,~~
(H_\tau d\bar{d}):\frac{U_R^{12}}{c_\beta}\simeq -\cot\beta,\notag\\
&(A_\tau \tau\tau): \frac{U_I^{22}}{c_\beta}\simeq -\tan\beta,~~
(A_\tau u\bar{u}):-\frac{U_I^{12}}{c_\beta}\simeq -\cot\beta,~~
(A_\tau d\bar{d}):\frac{U_I^{12}}{c_\beta}\simeq \cot\beta,\notag\\
&(H_\tau^+ \tau^-\nu): \frac{U_+^{22}}{c_\beta}\simeq -\tan\beta,~~
(H_\tau^+ d\bar{u}):\frac{U_+^{12}}{s_\beta}\simeq \cot\beta. 
\end{align}
It can be seen from the above expressions that the lepton (quark) couplings can be enhanced (suppressed) in the case with large $\tan\beta$. 
This feature in our model is very similar to that of the Type-X THDM.

\subsection{Current experimental constraints}
We here consider constraints from the current experimental data. 
As we discussed in the previous subsection, phenomenology of the lightest extra Higgs bosons ($H_\tau$, $A_\tau$ and $H_\tau^\pm$) 
can be regarded as that in the Type-X THDM, so that the current experimental 
bound can be applied to our model in the similar way as that in the Type-X THDM. 
We take into account the following constraints. 
\begin{enumerate}
\item At the LEP direct search experiment, 
masses for extra CP-odd, CP-even and charged Higgs bosons have been constrained from below by 93.4 GeV,  
92.8 GeV and 79.3 GeV, respectively, with the 95\% confidence level in supersymmetric (SUSY) models~\cite{PDG}. 
\item
From the $B$ physics experiments, the mass of charged Higgs bosons $m_{H^+}$ 
is strongly constrained in multi-doublet models. 
For example, from the $B\to X_s\gamma$ data, the lower limit of $m_{H^+}$ has been given by 295 GeV~\cite{Misiak} 
with the 95\% confidence level in the Type-II THDM with $\tan\beta \gtrsim 2$. 
However, this constraint turns out to be quite weak in the Type-I and Type-X THDMs;
{\it i.e.}, $m_{H^+}=100$ GeV with $\tan\beta\gtrsim 3$ is allowed with the 95\% confidence level as shown in Ref.~\cite{typeX,typeX2,Stal}.
The other processes including the $B$ meson such as $B\to \tau\nu$, $B\to D\tau\nu$, etc. give milder bounds 
compared to that from the $B\to X_s\gamma$ process\footnote{Recently, 
BaBar Collaboration has reported data 
on the ratios $\text{BR}(B \to D^{*} \tau \nu) / \text{BR}(B \to D^{*} \ell \nu)$ and $\text{BR}(B \to D \tau \nu) / \text{BR}(B \to D \ell \nu)$ ($\ell = e, \mu$) that deviate from the SM expectations by $2.7\sigma$ and $2.0\sigma$, respectively, and their combined deviation is $3.4\sigma$~\cite{Babar}.  
These deviations cannot be simultaneously explained by the contributions of the charged Higgs boson in the 
softly-broken $\mathbb{Z}_2$ symmetric THDMs. } in the Type-X THDM. 
\item At LHC, the ATLAS and the CMS Collaborations have reported the signal strength for a Higgs boson like 
particle with the mass of around 126 GeV. 
So far, five decay modes of the Higgs boson have been mainly analyzed, those are 
$h\to \gamma\gamma$, $h\to ZZ^*$, $h\to WW^*$, $h\to b\bar{b}$ and $h\to \tau\tau$. 
Their signal strengths are consistent with the prediction in the SM within the two-sigma level~\cite{Moriond_ATLAS,Moriond_CMS}. 
Thus, the parameter regions which give the SM-like limit explained in Eq.~(\ref{SMlike}) are also allowed in our model. 
\item
The extra neutral Higgs bosons in the minimal SUSY SM (MSSM); {\it i.e.}, the CP-even Higgs boson $H$ and 
the CP-odd Higgs boson $A$ have been searched by using the $\tau$ pair decay mode 
in the gluon fusion process $gg\to H/A\to \tau\tau$ and the bottom quark associated process 
$gg\to b\bar{b} H/A\to b\bar{b}\tau\tau$~\cite{MSSM_neutral_ATLAS,MSSM_neutral_CMS} by using the data collected at LHC with the collision energy 
to be 7 TeV. 
When the mass of the CP-odd Higgs boson $m_A$ is taken to be from 110 GeV to 150 GeV, 
the 95\% confidence level lower limit for $\tan\beta$ has been obtained to be about 10 in Ref.~\cite{MSSM_neutral_ATLAS}. 
Although the Higgs sector in the MSSM corresponds to the Type-II THDM, 
this constraint cannot be simply applied to the non-SUSY Type-II THDM due to SUSY relations. 
Understanding this, let us assume that the 95\% confidence level upper limit for 
the cross section of $gg\to b\bar{b} H/A\to b\bar{b}\tau\tau$ and $gg\to b\bar{b} H/A\to b\bar{b}\tau\tau$ 
is given by these cross sections calculated in the Type-II THDM with the case of $110<m_A(=m_H)<150\text{ GeV}$, $\sin(\beta-\alpha)=1$ and $\tan\beta=10$.
The cross section can be calculated by 
\begin{align}
\sigma_{95}^{\text{Type-II}} &= \sum_{\Phi^0=H,A}[\sigma_{gg\to \Phi^0}+\sigma_{gg\to b\bar{b}\Phi^0}]\times \text{BR}(\Phi^0\to \tau\tau)
\notag\\
&=
\sum_{\Phi^0=H,A}\left[\sigma_{gg\to h_\text{SM}}\frac{\Gamma_{\Phi^0\to gg}}{\Gamma_{h_{\text{SM}}\to gg}}+\sigma_{gg\to b\bar{b}h_\text{SM}} \tan^2\beta\right]\times \text{BR}(\Phi^0\to \tau\tau), \label{limit}
\end{align}
where each of $\sigma_{gg\to \Phi^0}$ ($\sigma_{gg\to h_{\text{SM}}}$) and $\sigma_{gg\to b\bar{b}\Phi^0}$ ($\sigma_{gg\to b\bar{b}h_{\text{SM}}}$)
is the cross sections of the $gg\to \Phi^0(h_{\text{SM}})$  and $gg\to b\bar{b}\Phi^0(b\bar{b}h_{\text{SM}})$ processes with 
$h_{\text{SM}}$ being the SM Higgs boson whose mass is taken to be the same as the mass of $\Phi^0$.  
In the Type-X THDM, $\text{BR}(\Phi^0\to \tau\tau)$ can be almost 100\% when $\tan\beta\gtrsim 2$. 
On the other hand, the quark couplings with $H$ and $A$ are proportional to $\cot\beta$, so that 
the cross section can be the maximal value at around $\tan\beta\simeq 2$~\cite{typeX}. 
In Tab.~\ref{tanb_lim}, the lower limit for the value of $\tan\beta$ in the Type-X THDM in the case with 
$\sin(\beta-\alpha)=1$ and $m_A=m_H=m_{\Phi^0}$ is listed, which is 
obtained by imposing the upper limit for the cross section given in Eq.~(\ref{limit}). 
We note that this constraint for $\tan\beta$ can be relaxed when the mass degeneracy in $A$ and $H$ is not assumed. 
\item
The charged Higgs boson $H^\pm$ in the MSSM has been searched from the top quark decay $t\to H^\pm b$ at LHC. 
In Refs.~\cite{MSSM_charged_ATLAS,MSSM_charged_CMS}, the excluded regions at the 95\% confidence level 
in the $\tan\beta$-$m_{H^+}$ plane are shown. 
However, in the Type-X THDM, the decay rate of $t\to H^\pm b$ is suppressed by the factor of $\cot^2\beta$~\cite{typeX,typeX2}, 
so that the constraint from the top decay becomes much weaker than that obtained in the MSSM.  

\end{enumerate}

\begin{table}[t]
\begin{center}
\begin{tabular}{c|c|c|c|c}\hline \hline 
$m_{\Phi^0}$ [GeV]& $\sigma_{gg\to h_{\text{SM}}}$ [pb]~\cite{Higgs_cross}&$\sigma_{gg\to b\bar{b}h_{\text{SM}}}$ [pb]&$\sigma_{95}^{\text{Type-II}}$ [pb]&$\tan\beta$ \\\hline
110&  19.8& 0.212 & 5.60 & 2.5\\\hline
120&  16.7& 0.155 & 4.00 & 2.9\\\hline
130&  14.2& 0.116 & 2.93 & 3.2\\\hline
140&  12.2& 0.089 & 2.20 & 3.6\\\hline
150&  10.6& 0.068 & 1.67 & 3.8\\\hline\hline
\end{tabular} 
\caption{Cross sections with the collision energy to be 7 TeV at LHC and the lower limit for $\tan\beta$ in the Type-X THDM. 
In the Second and third columns, the cross sections of $gg\to h_{\text{SM}}$ and $gg\to b\bar{b}h_{\text{SM}}$ processes in the SM are given in the case where the SM Higgs boson mass is taken to be $m_{\Phi^0}$. 
The former cross section is referred from~\cite{Higgs_cross}, and the latter one is calculated by using {\tt MadGraph5}~\cite{MG5}. 
In each $m_{\Phi^0}$, $\sigma_{95}^{\text{Type-II}}$ is calculated by using Eq.~(\ref{limit}) in the Type-II THDM with $\tan\beta=10$. 
The numbers displayed in the last column show the lower limit for $\tan\beta$ in the Type-X THDM by using $\sigma_{95}^{\text{Type-II}}$ at the 95\% confidence level. }
\label{tanb_lim}
\end{center}
\end{table}

According to the above constraints, we here give an example of the allowed parameter set as  
\begin{align}
&v_{\Delta \tau}=v_{\Delta \mu}=v_{\Delta e}=2.2~\text{GeV},~\mu =3 ~\text{TeV},~\tilde{\mu}=300~\text{GeV},~
\tilde{m}^2_{ \Phi}=-(100~\text{GeV})^2,\\
&\lambda_q= 0.19,~\lambda_\ell=5,~\lambda_{q\ell}=-0.035,~\lambda_\pm=-1,~\tan\beta =3.98. 
\end{align}
We then obtain the following outputs 
\begin{align}
&(m_h,m_{H_\tau},m_{H_\mu},m_{H_e})=(126~\text{GeV},172~\text{GeV},411~\text{GeV},6076~\text{GeV}),\\
&U_R=\left(
\begin{array}{cccc}
0.966 & -0.257 &  4.69\times 10^{-4} & 0\\
0.256&  0.964 & 0.0713& 2.71\times 10^{-4}\\
0.0187& 0.0688&  -0.997& 2.72\times 10^{-4}\\
7.43\times 10^{-5}& 2.80\times 10^{-4}&  -2.52\times 10^{-4}& -1.
\end{array}\right).
\end{align}
\begin{align}
&(m_{G^0},m_{A_\tau},m_{A_\mu},m_{A_e})=(0,110~\text{GeV},433~\text{GeV},6077~\text{GeV}),\\
&U_I=\left(
\begin{array}{cccc}
  0.969& 0.244& 0.000646& 0\\
  0.244& -0.968& 0.0568& 2.71\times 10^{-4}\\
  0.0145& -0.0550& -0.998& 2.72\times 10^{-4}\\
  6.70\times 10^{-5}& -2.77\times 10^{-4}& -2.56\times 10^{-4}& -1
\end{array}\right),\\
&(m_{G^+},m_{H_\tau^+},m_{H_\mu^+},m_{H_e^+})=(0,191~\text{GeV},455~\text{GeV},6079~\text{GeV}),\\
&U_+=\left(
\begin{array}{cccc}
  0.969&-0.244&0.000333& 0\\
  0.244& 0.968&0.0580&2.71\times 10^{-4}\\
  0.0145& 0.0562&-0.998&2.72\times 10^{-4}\\
  7.00\times 10^{-5} & 2.77\times 10^{-4}&-2.56\times 10^{-4}&-1
\end{array}\right).
\end{align}
As it can be seen the value of elements in $U_R$, this set is one of the realizations of the SM-like limit defined in Eq.~(\ref{SMlike}). 
The masses of the triplet-like Higgs bosons whose components are mainly from the triplet Higgs field ($\Delta_\tau$, $\Delta_\mu$ and 
$\Delta_e$)
including the doubly-charged Higgs bosons are around 2-3 TeV. 
These magnitudes are typical in our model, because the (squared) triplet-like Higgs boson masses are given like 
$(\mu+\tilde{\mu})v^2/v_\Delta$ as we discussed in the previous section, and $\mu,\tilde{\mu}$ ($v_\Delta$) 
have to be of order 1 TeV (1 GeV) to raise the $A_\tau$ mass. 

Finally, we comment on collider signatures of extra Higgs bosons. 
Although the existence of the doubly-charged Higgs bosons 
can be a clear signature of the model, their masses are too heavy to directly produce at LHC. 
Thus, we focus on the extra CP-even, CP-odd and singly-charged Higgs bosons. 
Phenomenology of these lightest Higgs bosons are similar to those 
in the Type-X THDM, and detailed studies of their collider signatures have been analyzed in Ref.~\cite{typeX,Yokoya} at LHC and the 
International Linear Collider, based on the $\tau$ specific nature of them. 
Therefore, we consider the collider signature of the second lightest extra Higgs bosons; namely, $A_\mu$, $H_\mu$ and $H_\mu^\pm$ 
at LHC. 
They can dominantly couple to the muon, because their magnitude are determined by $U_{R,I,+}^{33}$ whose values are almost unity; {\it e.g.}, 
in the parameter sets given in the above, we obtain $U_{R}^{33}=-0.997$, $U_{I}^{33}=-0.998$ and $U_{+}^{33}=-0.998$. 
When we neglect decay modes of a scalar to lighter two scalars
 such as $H_\mu \to h h$, the decay branching fractions of $H_\mu\to
 \mu^+\mu^-$, $A_\mu\to \mu^+\mu^-$ and $H_\mu^\pm \to \mu^\pm \nu$ are 
 almost 100\%.
 In such a case, the tetra-muon process $pp\to Z^* \to A_\mu H_\mu
 \to\mu^+\mu^-\mu^+\mu^- $,
 the tri-muon process $pp\to W^* \to A_\mu/H_\mu H_\mu^\pm
 \to\mu^+\mu^-\mu^\pm \nu$ and
 the di-muon process $pp\to Z^*/\gamma^* \to H_\mu^+ H_\mu^-
 \to\mu^+\mu^-\nu \nu$ can be signals of the model.
 At LHC with the collision energy of 14 TeV,
 the cross sections are 0.87 fb for the tetra-muons, 2.8 fb for the
 tri-muon and 0.68 fb for the di-muon processes in the case with the above
 parameter sets, which are obtained by using 
 {\tt CalcHEP3.4.2}~\cite{CalcHEP} and the
 {\tt CTEQ6L} parton distribution functions.
 We would like to emphasize that simultaneous observations of the
 signature expected in the Type-X THDM and multi-muon signatures from
 the muon specific Higgs bosons ($H_\mu$, $A_\mu$ and $H_\mu^\pm$) are
 important to test our model.

\vspace*{-4mm}

\section{Conclusions}

We have constructed a loop-induced neutrino mass model and analyzed Higgs phenomenologies with $T_7$ flavor symmetry in a renormalizable theory. In our model,
we have shown that observed neutrinos and their mixings can be generated through 
a three-loop level diagram (that derives their diagonal elements) in combination with 
the type-II seesaw (that derives their off-diagonal elements). 
Also we have analyzed the Higgs phenomenology
which can be reduced to that in the Type-X THDM, and found a benchmark point which is 
consistent with several constraints by the current experiments at such as LHC. 
Since the second lightest CP-odd, -even and charged Higgs bosons mainly 
decay into muons, our model is testable by observing multi-muon signatures.

\section*{Acknowledgments}
H.O. thanks to Prof. Eung-Jin Chun for fruitful discussion.
Y.K. thanks Korea Institute for Advanced Study for the travel support and local hospitality
during some parts of this work.
K.Y. was supported in part by the National Science Council of R.O.C. under Grant No. NSC-101-2811-M-008-014.

\begin{appendix}

\section{Details for the Higgs sector}

In this appendix, we give the detailed expressions for the 
Higgs potential, tadpole conditions and mass matrices for the Higgs bosons
in the flavor indices $(1,2,3)$, which correspond to $(\tau, e, \mu)$ in the main text.   

\subsection{Higgs potential}

The most general $T_7$ invariant Higgs potential is given as follows
\begin{align}
&V_{\text{Higgs}}
=
m_1^2 \Phi^\dagger_a \Phi_a 
+m_2^2 \Phi_q^\dag \Phi_q
+m_3^2  {\rm Tr} (\Delta^\dag_a\Delta_a) 
+ m_4^2 \eta^\dagger_a \eta_a 
\notag \\&
+\mu(\Phi_3^Ti\tau_2\Delta_1^\dag \Phi_3+\Phi_1^T i\tau_2\Delta_2^\dag \Phi_1+\Phi_2^T i\tau_2\Delta_3^\dag \Phi_2+\text{h.c.})
\notag\\&
+\mu_\eta (\eta^T_1i\tau_2 \Delta_2^\dag \eta_3+\eta^T_2i\tau_2 \Delta_3^\dag \eta_1
+\eta^T_3 i\tau_2\Delta_1^\dag \eta_2+\text{h.c.})
\notag \\&
+\lambda_1^{(1)} (\Phi^\dagger_a \Phi_a)^{2}
+\lambda_1^{(2)} (\omega^{2(a-1)}\Phi^\dagger_a \Phi_a) 
(\omega^{b-1}\Phi^\dagger_b \Phi_b) 
+\lambda_1^{(4)} (|\Phi^\dagger_1 \Phi_2|^2
+|\Phi^\dagger_2 \Phi_3|^2+|\Phi^\dagger_3 \Phi_1|^2)
\notag \\&
+\lambda_2^{(1)} (\eta^\dagger_a \eta_a)^{2}
+\lambda_2^{(2)} (\omega^{2(a-1)}\eta^\dagger_a \eta_a) 
(\omega^{b-1}\eta^\dagger_b \eta_b) 
+\lambda_2^{(4)} (|\eta^\dagger_1 \eta_2|^2
+|\eta^\dagger_2 \eta_3|^2+|\eta^\dagger_3 \eta_1|^2 )
\notag \\&
+\lambda_3^{(1)}(\Phi^\dagger_a \Phi_a)(\eta^\dagger_b \eta_b)
+\lambda_3^{(2)} [(\omega^{2(a-1)}\Phi^\dagger_a \Phi_a) 
(\omega^{b-1}\eta^\dagger_b \eta_b) +\text{h.c.}]
\notag \\&
+\lambda_3^{(4)}[ (\Phi^\dagger_1 \Phi_2)(\eta^\dagger_1 \eta_2)
+ (\Phi^\dagger_2 \Phi_3)(\eta^\dagger_2 \eta_3)
+ (\Phi^\dagger_3 \Phi_1)(\eta^\dagger_3 \eta_1)+\text{h.c.}]
\notag \\&
+\lambda_4^{(1)}(\Phi^\dagger_a \eta_b)(\eta^\dagger_b \Phi_a)
+\lambda_4^{(2)}[\omega^{2(a-1)}\omega^{b-1}
(\Phi^\dagger_a \eta_b)(\eta^\dagger_b \Phi_a)+\text{h.c.} ]
\notag \\&
+\lambda_4^{(4)} [ (\Phi^\dagger_2 \eta_1)(\eta^\dagger_2 \Phi_1)
+(\Phi^\dagger_3 \eta_2)(\eta^\dagger_3 \Phi_2)
+(\Phi^\dagger_1 \eta_3)(\eta^\dagger_1 \Phi_3)+\text{h.c.} ]
\notag \\&
+\lambda_5^{(1)} [  (\Phi^\dagger_1 \eta_2) (\Phi^\dagger_2 \eta_2)
+ (\Phi^\dagger_2 \eta_3) (\Phi^\dagger_3 \eta_3)
+ (\Phi^\dagger_3 \eta_1) (\Phi^\dagger_1 \eta_1)+\text{h.c.} ]
\notag \\&
+\lambda_5^{(2)} [(\Phi^\dagger_2 \eta_1) (\Phi^\dagger_2 \eta_2)
+(\Phi^\dagger_3 \eta_2) (\Phi^\dagger_3 \eta_3)
+  (\Phi^\dagger_1 \eta_3) (\Phi^\dagger_1 \eta_1)
 +\text{h.c.} ]
\notag \\&
+ \lambda_{6}^{(1)} [{\rm Tr} (\Delta_a^\dag\Delta_a ) ]^2
+ \lambda_{6}^{(2)}
 {\rm Tr} (\omega^{2(a-1)}\Delta_a^\dag\Delta_a) 
 {\rm Tr} (\omega^{b-1}\Delta_b^\dag\Delta_b) 
\notag \\&
+ \lambda_{6}^{(4)} [ {\rm Tr} (\Delta_2^\dag\Delta_1) {\rm Tr} (\Delta_1^\dag\Delta_2)
+{\rm Tr} (\Delta_3^\dag\Delta_2 ){\rm Tr} (\Delta_2^\dag\Delta_3)+
{\rm Tr} (\Delta_1^\dag\Delta_3 ){\rm Tr} (\Delta_3^\dag\Delta_1 )]
+\lambda_7{\rm Det} (\Delta_a^\dag\Delta_a )
\notag \\&
+\lambda_8^{(1)} (\Phi_a^\dag \Phi_a)(\Phi_q^\dag \Phi_q)
+\lambda_8^{(2)} (\Phi_a^\dag \Phi_q)(\Phi_q^\dag \Phi_a)
+\lambda_9^{(1)} (\eta_a^\dag \eta_a)(\Phi_q^\dag \Phi_q)
+\lambda_9^{(2)} (\eta_a^\dag \Phi_q)(\Phi_q^\dag \eta_a)
+\lambda_{10}(\Phi_q^\dag \Phi_q)^2\notag \\&
+ a_1^{(1)}  {\rm Tr} (\Delta_a^\dag\Delta_a )
 (\eta^\dagger_b \eta_b )
+ a_1^{(2)} [ {\rm Tr} (\omega^{2(a-1)}\Delta_a^\dag\Delta_a ) 
 (\omega^{b-1}\eta^{\dag}_b \eta_b )+\text{h.c.} ]
\notag \\&
+ a_1^{(4)} 
 [ {\rm Tr} (\Delta_1^\dag \Delta_2 )(\eta_1^\dag\eta_2)
+{\rm Tr} (\Delta_2^\dag\Delta_3 )(\eta_2^\dag\eta_3)+
{\rm Tr} (\Delta_3^\dag\Delta_1 )(\eta_3^\dag\eta_1)  +\text{h.c.}   ]
\notag \\&
+ a_2^{(1)}  {\rm Tr} (\Delta_a^\dag\cdot\Delta_a ) 
 (\eta^\dagger_b \cdot\eta_b )
+ a_2^{(2)} [ {\rm Tr}
 (\omega^{2(a-1)}\Delta_a^\dag\cdot\Delta_a ) 
 (\omega^{b-1}\eta^{\dag}_b\cdot \eta_b )+ \text{h.c.} ]
\notag \\&
+ a_2^{(4)} 
 [ {\rm Tr} (\Delta_1^\dag\cdot\Delta_2)
(\eta_1^\dag\cdot\eta_2)
+{\rm Tr} (\Delta_2^\dag\cdot\Delta_3 )(\eta_2^\dag\cdot\eta_3)+
{\rm Tr} (\Delta_3^\dag\cdot\Delta_1 )(\eta_3^\dag\cdot\eta_1)  +\text{h.c.}   ]
\notag \\&
+ b_1^{(1)}  {\rm Tr} (\Delta_a^\dag\Delta_a ) 
 (\Phi^\dagger_b \Phi_b )
+ b_1^{(2)} [ {\rm Tr}
 (\omega^{2(a-1)}\Delta_a^\dag\Delta_a ) 
 (\omega^{b-1}\Phi^{\dag}_b \Phi_b )+\text{h.c.} ]
\notag \\&
+ b_1^{(4)} 
[ {\rm Tr} (\Delta_1^\dag \Delta_2 )(\Phi_2^\dag\Phi_1)
+{\rm Tr} (\Delta_2^\dag\Delta_3 )(\Phi_3^\dag\Phi_2)+
{\rm Tr} (\Delta_3^\dag\Delta_1 )(\Phi_1^\dag\Phi_3)  +\text{h.c.}   ]
\notag \\&
+ b_2^{(1)}  {\rm Tr} (\Delta_a^\dag\cdot\Delta_a ) 
 (\Phi^\dagger_b \cdot\Phi_b )
+ b_2^{(2)} [ {\rm Tr}
 (\omega^{2(a-1)}\Delta_a^\dag\cdot\Delta_a )
 (\omega^{b-1}\Phi^{\dag}_b\cdot \Phi_b )+\text{h.c.} ]
\notag \\&
+ b_2^{(4)} [
  {\rm Tr} (\Delta_1^\dag\cdot\Delta_2 )(\Phi_2^\dag\cdot\Phi_1)
+{\rm Tr} (\Delta_2^\dag\cdot\Delta_3 )(\Phi_3^\dag\cdot\Phi_2)+
{\rm Tr} (\Delta_3^\dag\cdot\Delta_1)(\Phi_1^\dag\cdot\Phi_3)  +\text{h.c.}   ]
\notag \\&
+c_1  {\rm Tr} (\Delta_a^\dag \Delta_a )
 (\Phi^\dagger_q \Phi_q )
+ c_2 {\rm Tr} (\Delta_a^\dag\cdot\Delta_a )
 (\Phi^\dagger_q \cdot\Phi_q ), \label{pot_gen}
\end{align}
where $\cdot \equiv \tau_i$. 
When terms with indices $a$ and $b$ appear in the potential, 
they are summed over $a,b=1,2,3$.  
We give the correspondence between the dimensionless coupling constants defined in Eq.~(\ref{pot_gen}) and
those defined in the main text as
\begin{align}
&\lambda_1^{(1)}=\lambda_{\ell1},~\lambda_1^{(2)}=\lambda_{\ell2},~\lambda_1^{(4)}=\lambda_{\pm},~
\lambda_8^{(1)}=\lambda_{q\ell1},~\lambda_8^{(2)}=\lambda_{q\ell2},~\lambda_{10}=\lambda_q. 
\end{align}
In addition to this, 
we introduce the following soft breaking terms 
\begin{align}
V_{\text{soft}}&=
  \tilde m_{1a}^2 \Phi_a^\dag \Phi_a 
+\tilde m_{3a}^2 {\rm Tr} (\Delta^{\dag}_a \Delta_a )
+\tilde m_{4a}^2 \eta_a^{\dag} \eta_a\notag \\&
+\tilde m_{ \Phi}^2  ( \Phi_1^\dag \Phi_2+\Phi_2^\dag \Phi_3
+\Phi_3^\dag \Phi_1+\text{h.c.}   )
+\tilde m_{\eta}^2  ( \eta_1^{\dag} \eta_2+\eta_2^{\dag} \eta_3
+\eta_3^{\dag} \eta_1+\text{h.c.} )\notag \\&
+\tilde{\mu}(\Phi^T_q i\tau_2\Delta_a^{\dag}\Phi_q+\text{h.c.} ),
\label{soft}
\end{align}
which reduce the $T_7$ symmetry to $\mathbb{Z}_3$. 
Although the general soft breaking terms of the $T_7$ symmetry contain more mass 
terms\footnote
{Notice that Eq.~(\ref{soft}) is not general $\mathbb{Z}_3$ symmetric term. For instance, 
$[\Phi_q^{\dag}(\Phi_1+\Phi_2+\Phi_3)+\text{h.c.}]$ and $[\rm{Tr} (\Delta^\dag_1\Delta_2) +\rm{Tr} (\Delta^\dag_2\Delta_3)+
\rm{Tr} (\Delta^\dag_3\Delta_1)+\text{h.c.}]$ are allowed by the $\mathbb{Z}_3$ symmetry. 
}, we assume the minimal breaking of $T_7$ to $\mathbb{Z}_3$ that corresponds to 
cyclic permutation $\varphi_1 \to \varphi_2 \to \varphi_3 \to \varphi_1$ of each $T_7$ triplet. 

\subsection{Tadpole conditions}

Hereafter we define linear combinations of coupling constants as 
\begin{eqnarray*}
\lambda_1^{(124)}\equiv 2 \lambda_1^{(1)}-\lambda_1^{(2)}+\lambda_1^{(4)},~
 \lambda_8^{(1+2)}\equiv \lambda_8^{(1)}+\lambda_8^{(2)},~
 c_{1+2}\equiv c_1+c_2,~{\rm etc.}
\end{eqnarray*} 
Neglecting terms of ${\cal O}(v_{\Delta}^2)$, the tadpole conditions for each VEVs defined by Eqs. (\ref{vev-1}) and (\ref{vev-2}) are
written as follows
\begin{eqnarray*}
m_2^2&\simeq&-\lambda_{10}v_q^2
-\frac 12\lambda_8^{(1+2)}\left( v_1^2+v_2^2+v_3^2\right)
+\sqrt{2}\tilde{\mu }\left( v_{\Delta 1}+v_{\Delta 2}+v_{\Delta 3}\right),\\
\tilde{m}_{11}^2+m_1^2&\simeq&
-\frac{v_2+v_3}{v_1}\tilde{m}_{\Phi}^2 -\frac 12\left[ \lambda_8^{(1+2)}v_q^2
+2 \lambda_1^{(1+2)}v_1^2
+\lambda_1^{(124)}
\left( v_2^2+v_3^2\right)\right]
+{\sqrt{2}}\mu  v_{\Delta 2},\\
\tilde{m}_{12}^2+m_1^2&\simeq&
-\frac{v_3+v_1}{v_2}\tilde{m}_{\Phi}^2-\frac 12\left[\lambda_8^{(1+2)}v_q^2
+2 \lambda_1^{(1+2)}v_2^2
+ \lambda_1^{(124)}
\left( v_3^2+v_2^2\right)\right]
+{\sqrt{2}}\mu  v_{\Delta 3},\\
\tilde{m}_{13}^2+m_1^2&\simeq&
-\frac{v_1+v_2}{v_3}\tilde{m}_{\Phi}^2-\frac 12\left[\lambda_8^{(1+2)}v_q^2
+2 \lambda_1^{(1+2)}v_3^2
+\lambda_1^{(124)}
\left( v_1^2+v_2^2\right)\right]
+{\sqrt{2}}\mu  v_{\Delta 1},\\
\tilde{m}_{31}^2+m_3^2&\simeq&
\frac{\sqrt{2}(\mu  v_3^2+\tilde{\mu } v_q^2)}{2v_{\Delta 1}}\\
&-&\frac 12\left[ c_{1+2}v_q^2
+ \left( b_{1+2}^{(1)}+2 b_{1+2}^{(2)}\right)v_1^2
+\left( b_{1+2}^{(1)}-b_{1+2}^{(2)}\right)\left(v_2^2+v_3^2\right)
+b_{1+2}^{(4)}v_1\frac{v_2 v_{\Delta 2}+v_3 v_{\Delta 3}}{v_{\Delta 1}}
\right],\nn\\
\tilde{m}_{32}^2+m_3^2&\simeq&
\frac{\sqrt{2}(\mu  v_1^2+\tilde{\mu } v_q^2)}{2v_{\Delta 2}}\\
&-&\frac 12\left[ c_{1+2}v_q^2
+ \left( b_{1+2}^{(1)}+2 b_{1+2}^{(2)}\right)v_2^2
+\left( b_{1+2}^{(1)}-b_{1+2}^{(2)}\right)\left(v_3^2+v_1^2\right)
+b_{1+2}^{(4)}v_2\frac{v_3 v_{\Delta 3}+v_1 v_{\Delta 1}}{v_{\Delta 2}}
\right],\nn\\
\tilde{m}_{33}^2+m_3^2&\simeq&
\frac{\sqrt{2}(\mu  v_2^2+\tilde{\mu } v_q^2)}{2v_{\Delta 3}}\\
&-&\frac 12\left[ c_{1+2}v_q^2
+ \left( b_{1+2}^{(1)}+2 b_{1+2}^{(2)}\right)v_3^2
+\left( b_{1+2}^{(1)}-b_{1+2}^{(2)}\right)\left(v_1^2+v_2^2\right)
+b_{1+2}^{(4)}v_3\frac{v_1 v_{\Delta 1}+v_2 v_{\Delta 2}}{v_{\Delta 3}}
\right].\nn
\label{tad3}
\end{eqnarray*}

\subsection{Scalar mass matrices}\label{a}

We write down the explicit form of scalar mass matrices.  
Although $\Phi$ and $\Delta$ mix with each other except 
the doubly charged components $\Delta^{\pm \pm}$, 
the inert doublets $\eta$ do not. Again we neglect terms of 
${\cal O}(v_{\Delta}^2)$. The symbol $(i,j)$ represents the $(i,j)$ element of 
mass matrix. 
\\
{\underline{$\phi^{0}_{(1,2,3,q)R},~\Delta^{0}_{(1,2,3)R}$}}\\
The elements for the CP-even scalar states are calculated as 
\begin{eqnarray*}
(1,1)&\simeq&
-\tilde{m}_{\Phi}^2 \frac{v_2+v_3}{v_1}+2\lambda_1^{(1+2)}v_1^2,~~
(2,2)\simeq
-\tilde{m}_{\Phi}^2 \frac{v_3+v_1}{v_2}+2\lambda_1^{(1+2)}v_2^2,\\
(3,3)&\simeq&
-\tilde{m}_{\Phi}^2 \frac{v_1+v_2}{v_3}+2\lambda_1^{(1+2)}v_3^2,~~
(4,4)=2 \lambda_{10}v_q^2,\\
(1,2)&\simeq&
\tilde{m}_{\Phi}^2
+\lambda_1^{(124)}v_1 v_2,~~
(1,3)\simeq
\tilde{m}_{\Phi}^2
+\lambda_1^{(124)}v_1 v_3,~~
(2,3)\simeq
\tilde{m}_{\Phi}^2
+\lambda_1^{(124)}v_2 v_3,\\
(1,4)&=&
\lambda_8^{(1+2)}v_1 v_q,~~~~~~~~~~
(2,4)=
\lambda_8^{(1+2)}v_2 v_q,~~~~~~~~~~
(3,4)=
\lambda_8^{(1+2)}v_3 v_q,\\
(5,5)&\simeq&
\frac{\sqrt{2}(\mu  v_3^2+\tilde{\mu } v_q^2)}{2 v_{\Delta 1}}
-\frac 12 b_{1+2}^{(4)}v_1 \frac{v_2 v_{\Delta 2}+v_3 v_{\Delta 3}}{v_{\Delta 1}},\\
(6,6)&\simeq&
\frac{\sqrt{2}(\mu  v_1^2+\tilde{\mu } v_q^2)}{2 v_{\Delta 2}}
-\frac 12 b_{1+2}^{(4)}v_2 \frac{v_1 v_{\Delta 1}+v_3 v_{\Delta 3}}{v_{\Delta 2}} ,\\
(7,7)&\simeq&
\frac{\sqrt{2}(\mu  v_2^2+\tilde{\mu } v_q^2)}{2 v_{\Delta 3}}
-\frac 12 b_{1+2}^{(4)}v_3 \frac{v_1 v_{\Delta 1}+v_2 v_{\Delta 2}}{v_{\Delta 3}} ,\\
(5,6)&\simeq&
\frac 12 b_{1+2}^{(4)}v_1 v_2,~~
(5,7)\simeq
\frac 12 b_{1+2}^{(4)}v_1 v_3,~~
(6,7)\simeq
\frac 12 b_{1+2}^{(4)}v_2 v_3,\\
(1,5)&\simeq&
\left( b_{1+2}^{(1)}+2 b_{1+2}^{(2)}\right)v_1 v_{\Delta 1}
+\frac 12 b_{1+2}^{(4)}\left( v_2 v_{\Delta 2}+v_3 v_{\Delta 3}\right),\\
(1,6)&\simeq&
-\sqrt{2} \mu  v_1 +\left( b_{1+2}^{(1)}-b_{1+2}^{(2)}\right)v_1 v_{\Delta 2}
+\frac 12 b_{1+2}^{(4)}v_2 v_{\Delta 1},\\
(1,7)&\simeq&
\left( b_{1+2}^{(1)}-b_{1+2}^{(2)}\right)v_1 v_{\Delta 3}
+\frac 12 b_{1+2}^{(4)}v_3 v_{\Delta 1},\\
(2,5)&\simeq&
\left( b_{1+2}^{(1)}-b_{1+2}^{(2)}\right)v_2 v_{\Delta 1}
+\frac 12 b_{1+2}^{(4)}v_1 v_{\Delta 2},\\
(2,6)&\simeq&
\left( b_{1+2}^{(1)}+2 b_{1+2}^{(2)}\right)v_2 v_{\Delta 2}
+\frac 12 b_{1+2}^{(4)}\left( v_1 v_{\Delta 1}+v_3 v_{\Delta 3}\right),\\
(2,7)&\simeq&
-\sqrt{2} \mu  v_2
+\left( b_{1+2}^{(1)}-b_{1+2}^{(2)}\right)v_2 v_{\Delta 3}
+\frac 12 b_{1+2}^{(4)}v_3 v_{\Delta 2},\\
(3,5)&\simeq&
-\sqrt{2} \mu  v_3
+\left( b_{1+2}^{(1)}-b_{1+2}^{(2)}\right)v_3 v_{\Delta 1}
+\frac 12 b_{1+2}^{(4)}v_1 v_{\Delta 3},\\
(3,6)&\simeq&
+\left( b_{1+2}^{(1)}-b_{1+2}^{(2)}\right)v_3 v_{\Delta 2}
+\frac 12 b_{1+2}^{(4)}v_2 v_{\Delta 3},\\
(3,7)&\simeq&
\left( b_{1+2}^{(1)}+2 b_{1+2}^{(2)}\right)v_3 v_{\Delta 3}
+\frac 12 b_{1+2}^{(4)}\left( v_1 v_{\Delta 1}+v_2 v_{\Delta 2}\right),\\
(4,5)&=&
v_q\left(c_{1+2} v_{\Delta 1}-\sqrt{2}\tilde{\mu }\right),
(4,6)=
v_q\left(c_{1+2} v_{\Delta 2}-\sqrt{2}\tilde{\mu }\right),
(4,7)=
v_q\left(c_{1+2} v_{\Delta 3}-\sqrt{2}\tilde{\mu }\right). 
\end{eqnarray*}

{\underline{$\phi^{0}_{(1,2,3,q)I},~\Delta^{0}_{(1,2,3)I}$}}\\
The mass matrix for the CP-odd scalar states has a zero eigenvalue which corresponds to 
NG boson eaten by the $Z$ boson. Each element is calculated by 
\begin{eqnarray*}
(1,1)&\simeq&
-\frac{v_2+v_3}{v_1}\tilde{m}_{\Phi}^2
+2\sqrt{2}\mu  v_{\Delta 2},~~
(2,2)\simeq
-\frac{v_3+v_1}{v_2}\tilde{m}_{\Phi}^2
+2\sqrt{2}\mu  v_{\Delta 3},\\
(3,3)&\simeq&
-\frac{v_1+v_2}{v_3}\tilde{m}_{\Phi}^2
+2\sqrt{2}\mu  v_{\Delta 1},~~
(4,4)=2\sqrt{2}\tilde{\mu } \left(v_{\Delta 1}+v_{\Delta 2}+v_{\Delta 3} \right),\\
(1,2)&\simeq&(1,3)\simeq(2,3)\simeq \tilde{m}_{\Phi}^2,~~
(1,4)=(2,4)=(3,4)=0,\\
(5,5)&\simeq&
\frac{\sqrt{2}(\mu  v_3^2+\tilde{\mu } v_q^2)}{2v_{\Delta 1}}
-\frac 12 b_{1+2}^{(4)}v_1\frac{v_2 v_{\Delta 2}+v_3 v_{\Delta 3}}{v_{\Delta 1} } ,\\
(6,6)&\simeq&
\frac{\sqrt{2}(\mu  v_1^2+\tilde{\mu } v_q^2)}{2v_{\Delta 2}}
-\frac 12 b_{1+2}^{(4)}v_2\frac{v_3 v_{\Delta 3}+v_1 v_{\Delta 1}}{v_{\Delta 2} },\\
(7,7)&\simeq&
\frac{\sqrt{2}(\mu  v_2^2+\tilde{\mu } v_q^2)}{2v_{\Delta 3}}
-\frac 12 b_{1+2}^{(4)}v_3\frac{v_1 v_{\Delta 1}+v_2 v_{\Delta 2}}{v_{\Delta 3} },\\  
(5,6)&\simeq&\frac 12 b_{1+2}^{(4)}v_1 v_2,~~
(5,7)\simeq\frac 12 b_{1+2}^{(4)}v_1 v_3,~~
(6,7)\simeq\frac 12 b_{1+2}^{(4)}v_2 v_3,\\
(1,5)&\simeq&-\frac 12 b_{1+2}^{(4)}\left( v_2 v_{\Delta 2}+v_3 v_{\Delta 3}\right),
(1,6)\simeq-\sqrt{2}\mu  v_1
+\frac 12 b_{1+2}^{(4)}v_2 v_{\Delta 1},
(1,7)\simeq\frac 12 b_{1+2}^{(4)}v_3 v_{\Delta 1},\\
(2,5)&\simeq&\frac 12 b_{1+2}^{(4)}v_1 v_{\Delta 2},
(2,6)\simeq -\frac 12 b_{1+2}^{(4)}\left(v_3 v_{\Delta 3}+v_1 v_{\Delta 1} \right),
(2,7)\simeq -\sqrt{2}\mu  v_2+\frac 12 b_{1+2}^{(4)}v_3 v_{\Delta 2},\\
(3,5)&\simeq&-\sqrt{2}\mu  v_3+\frac 12 b_{1+2}^{(4)}v_1 v_{\Delta 3},
(3,6)\simeq \frac 12 b_{1+2}^{(4)}v_2 v_{\Delta 3},
(3,7)\simeq -\frac 12 b_{1+2}^{(4)}\left(v_1 v_{\Delta 1}+v_2 v_{\Delta 2} \right),\\
(4,5)&=&(4,6)=(4,7)=-\sqrt{2}\tilde{\mu } v_q. 
\end{eqnarray*}
{\underline{$\phi^{\pm}_{(1,2,3,q)},~\Delta^{\pm}_{(1,2,3)}$}}\\
Similar to the mass matrix for the CP-odd scalar states, 
that for the singly-charged scalar states has a zero eigenvalue which corresponds to 
NG boson eaten by the $W$ boson. Each element is calculated by 
\begin{eqnarray*}
(1,1)&\simeq&-\frac{v_2+v_3}{v_1}\tilde{m}_{\Phi}^2
-\frac 12 \left[ \lambda_1^{(4)}\left( v_2^2+v_3^2\right)
+\lambda_8^{(2)}v_q^2-2\sqrt{2}\mu  v_{\Delta 2}\right],\\
(2,2)&\simeq&-\frac{v_3+v_1}{v_2}\tilde{m}_{\Phi}^2
-\frac 12 \left[ \lambda_1^{(4)}\left( v_3^2+v_1^2\right)
+\lambda_8^{(2)}v_q^2-2\sqrt{2}\mu  v_{\Delta 3}\right],\\
(3,3)&\simeq&-\frac{v_1+v_2}{v_3}\tilde{m}_{\Phi}^2
-\frac 12 \left[ \lambda_1^{(4)}\left( v_1^2+v_2^2\right)
+\lambda_8^{(2)}v_q^2-2\sqrt{2}\mu  v_{\Delta 1}\right],\\
(4,4)&\simeq&-\frac 12 \lambda_8^{(2)}\left(v_1^2+v_2^2+v_3^2 \right)
+\sqrt{2}\tilde{\mu } \left( v_{\Delta 1}+v_{\Delta 2}+v_{\Delta 3}\right),\\
(1,2)&\simeq&\tilde{m}_{\Phi}^2+\frac 12 \lambda_1^{(4)}v_1 v_2,~~
(1,3)\simeq\tilde{m}_{\Phi}^2+\frac 12 \lambda_1^{(4)}v_3 v_1,~~
(2,3)\simeq\tilde{m}_{\Phi}^2+\frac 12 \lambda_1^{(4)}v_2 v_3,\\
(1,4)&=&\frac 12 \lambda_8^{(2)}v_1 v_q,~~~~~~~~~~~
(2,4)=\frac 12 \lambda_8^{(2)}v_2 v_q,~~~~~~~~~~~
(3,4)=\frac 12 \lambda_8^{(2)}v_3 v_q,\\
(5,5)&\simeq&
\frac{\sqrt{2}(\mu  v_3^2+\tilde{\mu } v_q^2)}{2 v_{\Delta 1}}\nn\\
&-&\frac 12 \left[ \left( b_2^{(1)}+2b_2^{(2)}\right)v_1^2
+b_2^{(1-2)}\left(v_2^2+v_3^2 \right)
+b_{1+2}^{(4)}\frac{v_1}{v_{\Delta 1}}\left( v_2 v_{\Delta 2}+v_3 v_{\Delta 3}\right)
+c_2 v_q^2\right],\\
(6,6)&\simeq&
\frac{\sqrt{2}(\mu  v_1^2+\tilde{\mu } v_q^2)}{2 v_{\Delta 2}}\nn\\
&-&\frac 12 \left[ \left( b_2^{(1)}+2b_2^{(2)}\right)v_2^2
+b_2^{(1-2)}\left(v_3^2+v_1^2 \right)
+b_{1+2}^{(4)}\frac{v_2}{v_{\Delta 2}}\left( v_3 v_{\Delta 3}+v_1 v_{\Delta 1}\right)
+c_2 v_q^2\right],\\
(7,7)&\simeq&
\frac{\sqrt{2}(\mu  v_2^2+\tilde{\mu } v_q^2)}{2 v_{\Delta 3}}\nn\\
&-&\frac 12 \left[ \left( b_2^{(1)}+2b_2^{(2)}\right)v_3^2
+b_2^{(1-2)}\left(v_1^2+v_2^2 \right)
+b_{1+2}^{(4)}\frac{v_3}{v_{\Delta 3}}\left( v_1 v_{\Delta 1}+v_2 v_{\Delta 2}\right)
+c_2 v_q^2\right],\\
(5,6)&\simeq&\frac 12 b_1^{(4)}v_1 v_2,~~~~~
(5,7)\simeq\frac 12 b_1^{(4)}v_3 v_1,~~~~~
(6,7)\simeq\frac 12 b_1^{(4)}v_2 v_3,\\
(1,5)&=&\frac{1}{\sqrt{2}}\left( b_2^{(1)}+2 b_2^{(2)}\right)v_1 v_{\Delta 1},~~
(1,6)=-\mu  v_1+\frac{1}{\sqrt{2}}\left[ b_2^{(1-2)}v_1 v_{\Delta 2}
+b_2^{(4)}v_2 v_{\Delta 1}\right],\\
(1,7)&=&\frac{1}{\sqrt{2}}\left[ b_2^{(1-2)}v_1 v_{\Delta 3}
+b_2^{(4)}v_3 v_{\Delta 1}\right],\\
(2,5)&=&\frac{1}{\sqrt{2}}\left[ b_2^{(1-2)}v_2 v_{\Delta 1}
+b_2^{(4)}v_1 v_{\Delta 2}\right],~~
(2,6)=\frac{1}{\sqrt{2}}\left( b_2^{(1)}+2 b_2^{(2)}\right)v_2 v_{\Delta 2},\\
(2,7)&=&-\mu  v_2+\frac{1}{\sqrt{2}}\left[ b_2^{(1-2)}v_2 v_{\Delta 3}
+b_2^{(4)}v_3 v_{\Delta 2}\right],\\
(3,5)&=&-\mu  v_3+\frac{1}{\sqrt{2}}\left[ b_2^{(1-2)}v_3 v_{\Delta 1}
+b_2^{(4)}v_1 v_{\Delta 3}\right],~~
(3,6)=\frac{1}{\sqrt{2}}\left[ b_2^{(1-2)}v_3 v_{\Delta 2}
+b_2^{(4)}v_2 v_{\Delta 3}\right],\\
(3,7)&=&\frac{1}{\sqrt{2}}\left( b_2^{(1)}+2 b_2^{(2)}\right)v_3 v_{\Delta 3},\\
(4,5)&=&v_q\left(\frac{1}{\sqrt{2}}c_2 v_{\Delta 1}-\tilde{\mu }\right),~
(4,6)=v_q\left(\frac{1}{\sqrt{2}}c_2 v_{\Delta 2}-\tilde{\mu }\right),~
(4,7)=v_q\left(\frac{1}{\sqrt{2}}c_2 v_{\Delta 3}-\tilde{\mu }\right). 
\end{eqnarray*}
{\underline{$\Delta^{\pm \pm}_{(1,2,3)}$}}\\
The elements of the mass matrix for the doubly-charged scalar states which are purely from the triplet Higgs fields 
are given by 
\begin{eqnarray*}
(1,1)&\simeq&
\frac{\sqrt{2}(\mu  v_3^2+\tilde{\mu } v_q^2)}{2 v_{\Delta 1}}\nn\\
&-&\left[\left( b_2^{(1)}+2b_2^{(2)}\right)v_1^2
+b_2^{(1-2)}\left( v_2^2+v_3^2\right) 
+\frac 12 b_{1+2}^{(4)}\frac{v_1}{v_{\Delta 1}}\left(v_2 v_{\Delta 2}+v_3 v_{\Delta 3}
 \right)\right],\\
(2,2)&\simeq&
\frac{\sqrt{2}(\mu  v_1^2+\tilde{\mu } v_q^2)}{2 v_{\Delta 2}}\nn\\
&-&\left[\left( b_2^{(1)}+2b_2^{(2)}\right)v_2^2
+b_2^{(1-2)}\left( v_3^2+v_1^2\right) 
+\frac 12 b_{1+2}^{(4)}\frac{v_2}{v_{\Delta 2}}\left(v_3 v_{\Delta 3}+v_1 v_{\Delta 1}
 \right)\right],\\
 (3,3)&\simeq&
\frac{\sqrt{2}(\mu  v_2^2+\tilde{\mu } v_q^2)}{2 v_{\Delta 3}}\nn\\
&-&\left[\left( b_2^{(1)}+2b_2^{(2)}\right)v_3^2
+b_2^{(1-2)}\left( v_1^2+v_2^2\right) 
+\frac 12 b_{1+2}^{(4)}\frac{v_3}{v_{\Delta 3}}\left(v_1 v_{\Delta 1}+v_2 v_{\Delta 2}
 \right)\right],\\
 (1,2)&\simeq&\frac 12 b_{1-2}^{(4)}v_1 v_2,~~~~~
 (1,3)\simeq\frac 12 b_{1-2}^{(4)}v_3 v_1,~~~~~
 (2,3)\simeq\frac 12 b_{1-2}^{(4)}v_2 v_3. 
\end{eqnarray*}
{\underline{$\eta^0_{(1,2,3)R}$}}\\
Since there are no tadpole conditions for $\eta_a$, 
the mass parameter $m_4^2$ is not vanished. 
The elements are given with $m_{4a}^2=m_4^2+\tilde m_{4a}^2$ by
\begin{eqnarray*}
(1,1)&\simeq&m_{41}^2
+\frac 12 \left[ \lambda_{3+4}^{(1)}\left( v_1^2+v_2^2+v_3^2\right)
+\lambda_{3+4}^{(2)}\left( 2 v_1^2-v_2^2-v_3^2\right)
+2 \lambda_5^{(1)}v_3 v_1 
+\lambda_9^{(1+2)}v_q^2
\right],\\
(2,2)&\simeq&m_{42}^2
+\frac 12 \left[ \lambda_{3+4}^{(1)}\left( v_1^2+v_2^2+v_3^2\right)
+\lambda_{3+4}^{(2)}\left( -v_1^2+2v_2^2-v_3^2\right)
+2 \lambda_5^{(1)}v_1 v_2 
+\lambda_9^{(1+2)}v_q^2
\right],\\
(3,3)&\simeq&m_{43}^2
+\frac 12 \left[ \lambda_{3+4}^{(1)}\left( v_1^2+v_2^2+v_3^2\right)
+\lambda_{3+4}^{(2)}\left( -v_1^2-v_2^2+2v_3^2\right)
+2 \lambda_5^{(1)}v_2 v_3 
+\lambda_9^{(1+2)}v_q^2
\right],\\
(1,2)&\simeq&\tilde m_{\eta}^2
+\frac 12\left[ \lambda_{3+4}^{(4)}v_1 v_2
+\lambda_5^{(2)}v_2^2
-\sqrt{2}\mu_\eta  v_{\Delta 3}\right],\\
(1,3)&\simeq&\tilde m_{\eta}^2
+\frac 12\left[ \lambda_{3+4}^{(4)}v_3 v_1
+\lambda_5^{(2)}v_1^2
-\sqrt{2}\mu_\eta  v_{\Delta 2}\right],\\
(2,3)&\simeq&\tilde m_{\eta}^2
+\frac 12\left[ \lambda_{3+4}^{(4)}v_2 v_3
+\lambda_5^{(2)}v_3^2
-\sqrt{2}\mu_\eta  v_{\Delta 1}\right]. 
\end{eqnarray*}
{\underline{$\eta^0_{(1,2,3)I}$}}\\
The mass matrix for CP-odd components of $\eta^0$ is 
given by replacing 
$
\lambda_5^{(1,2)} \to -\lambda_5^{(1,2)}$ and 
$\mu_\eta  \to -\mu_\eta \nn,$
in the mass matrix of CP-even components given above. \\
{\underline{$\eta^{\pm}_{(1,2,3)}$}}\\
The elements for the singly-charged component of $\eta_a$ are given by 
\begin{eqnarray*}
(1,1)&\simeq&
m_{41}^2+\frac 12 \left[\lambda_3^{(1)}\left( v_1^2+v_2^2+v_3^2\right)
+\lambda_3^{(2)}\left( 2 v_1^2-v_2^2-v_3^2\right)
+\lambda_9^{(1)}v_q^2\right],\\
(2,2)&\simeq&
m_{42}^2+\frac 12 \left[\lambda_3^{(1)}\left( v_1^2+v_2^2+v_3^2\right)
+\lambda_3^{(2)}\left( - v_1^2+2v_2^2-v_3^2\right)
+\lambda_9^{(1)}v_q^2\right],\\
(3,3)&\simeq&
m_{43}^2+\frac 12 \left[\lambda_3^{(1)}\left( v_1^2+v_2^2+v_3^2\right)
+\lambda_3^{(2)}\left( -v_1^2-v_2^2+2v_3^2\right)
+\lambda_9^{(1)}v_q^2\right],\\
(1,2)&\simeq&\tilde m_{\eta}^2+\frac 12 \lambda_3^{(4)}v_1 v_2,~~
(1,3)\simeq\tilde m_{\eta}^2+\frac 12 \lambda_3^{(4)}v_3 v_1,~~
(2,3)\simeq\tilde m_{\eta}^2+\frac 12 \lambda_3^{(4)}v_2 v_3.
\end{eqnarray*}
%

\section{Three-loop Neutrino mass formula}\label{three-neutrino-mass}

Here we give the diagonal elements of neutrino mass matrix through three-loop level diagram 
depicted in Fig.\ref{three-neut} in the original flavor basis. 
Defining $\phi_{\beta,\gamma}^+$ to be singly-charged bosons, it is written as  
\begin{eqnarray*}
\left(M_\nu\right)_{ii}^{(3)}&=& 
-\frac{4}{(16 \pi^2)^3}y_{\Delta}^3 \mu_\eta ^3 
\sum_{j,k,\ell=1}^3\sum_{\alpha=1}^3\sum_{\beta,\gamma=1}^7 |U|^2
\frac{1}{m_{\phi^+_{\beta}}^2-m_{e_{i+2}}^2}
\frac{1}{m_{\phi^+_{\gamma}}^2-m_{e_{i+1}}^2} I,
\end{eqnarray*}
where
\begin{eqnarray*}
I&=&\int_0^1 dx dy dz \delta (x+y+z-1) 
\int_0^1 d s d t d u \delta(s+t+u-1)
\left(1-t-\frac{y}{1-z}u \right)\frac{1}{z(1-z)A}\nn\\
&\times&
\sum_{a=\phi^+_{\beta},e_{i+2}}\sum_{b=\phi^+_{\gamma},e_{i+1}}
 (-1)^{a,b}
\left[ \left( V_R\right)_{i\ell}^2 \Delta^{ab}_{R,i\ell} \ln  \Delta^{ab}_{R,i\ell} 
-\left( V_I\right)_{i\ell}^2 \Delta^{ab}_{I,i\ell} \ln  \Delta^{ab}_{I,i\ell}\right], \\
\Delta^{ab}_{R(I),i\ell}&=&\frac 12 m_a^2-\frac{1}{2A}
\left[ s m^2_{\Delta^{++}_{\alpha}}+t m_b^2
+\frac{u}{z(1-z)}\left( x m^2_{\eta^+_j}+y m^2_{\eta_{{R(I)}\ell}}
+zm^2_{\eta^+_{k}}\right)\right], \\
A&=&\left( t+\frac{y}{1-z}u\right)^2-t
-\frac{y(1-y)}{z(1-z)}u, \\
U&=&(U_{+}')_{5+i,\beta}(U_{+}')_{6+i,\gamma}(U_{\Delta})_{i,\alpha}
(V_+)_{i+1,k}(V_+)_{i+2,j}, 
\end{eqnarray*}
and
\begin{eqnarray*}
(-1)^{a,b}=&&+1 ~{\rm for}~ (a,b)=(\phi_{\beta}^+,\phi_{\gamma}^+),~
(e_{i+2},e_{i+1}),\\
&&-1 ~{\rm for}~ (a,b)=(\phi_{\beta}^+,e_{i+1}),~(e_{i+2},\phi_{\gamma}^+).
\end{eqnarray*}
The matrices $U_{\Delta}$, $V_{R(I)}$ and $V_+$ are diagonalization matrix of 
$\Delta^{\pm\pm}$, $\eta_{R(I)} $ and $\eta^{\pm}$, respectively. 
$U_{+}'$ is $7 \times 7$ diagonalization matrix of singly-charged bosons.

\section{Other loops of Neutrino mass}\label{minor-cont}
\begin{figure}[t]
\begin{center}
 \includegraphics[width = 100mm]{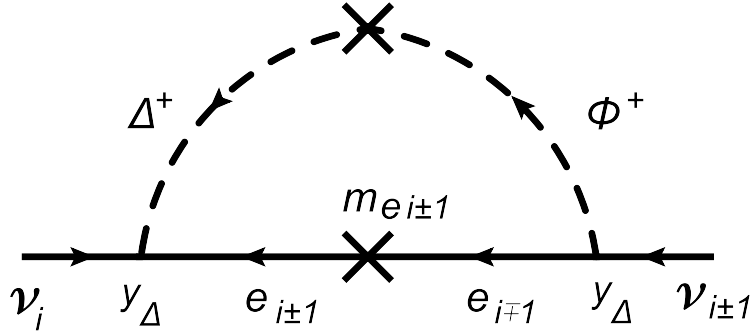}
   \caption{Neutrino mass generation via one-loop radiative seesaw. 
   There exist similar diagram with replacing $e \to \nu,~\Delta^+ \to \Delta^0,~
   \Phi^+ \to \Phi^0$. }
   \label{one-loop}
\end{center}
\end{figure}
\begin{figure}[t]
\begin{center}
 \includegraphics[width = 150mm]{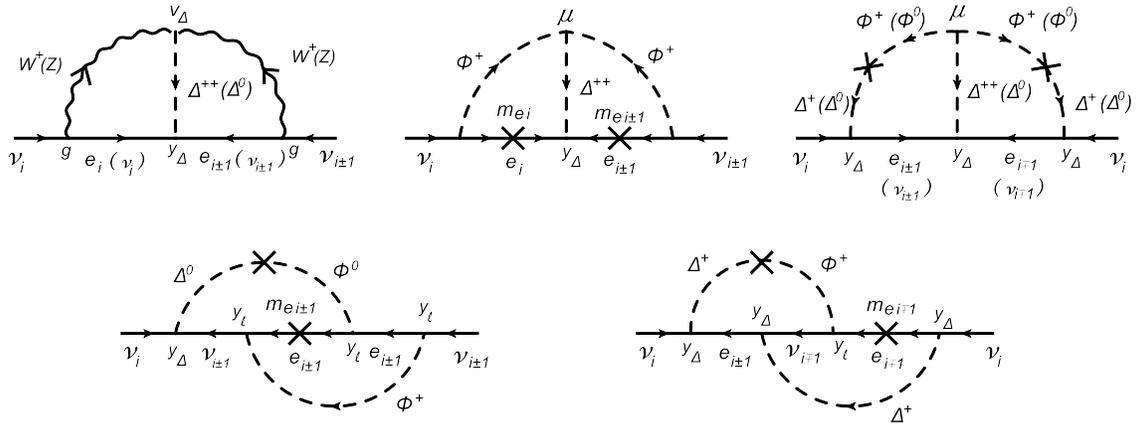}
   \caption{Zee-Babu like neutrino mass  via two-loop radiative seesaw \cite{zee-babu}. }
   \label{two-loop-1}
\end{center}
\end{figure}
\begin{figure}[t]
\begin{center}
 \includegraphics[width = 150mm]{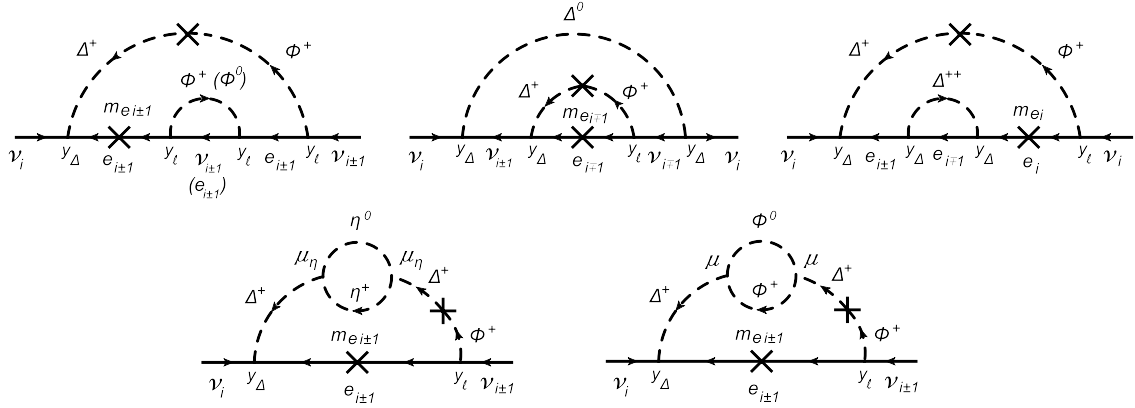}
   \caption{Rainbow-(lower panel) \cite{Kajiyama:2013zla} and 
   ring-(upper panel) \cite{Aoki:2013gzs} like  neutrino mass  via two-loop radiative seesaw. }
   \label{two-loop-2}
\end{center}
\end{figure}
\begin{figure}[t]
\begin{center}
 \includegraphics[width = 100mm]{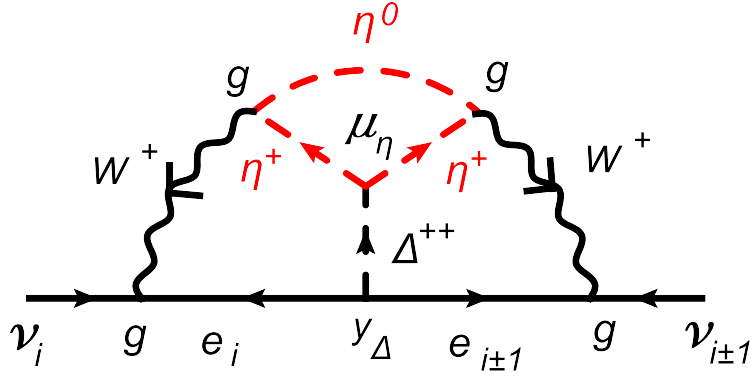}
   \caption{Cocktail like  Neutrino mass  via three-loop radiative seesaw \cite{Gustafsson}. We restrict ourselves the representative diagram.}
   \label{three-loop}
\end{center}
\end{figure}

In general, there exist several diagrams even up to two-loop level which generate 
neutrino masses of our model. Representative diagrams are shown in Figs.\ref{one-loop}-\ref{three-loop}.
Throughout these figures, the same flavor indices of the two external neutrinos give the diagonal elements of neutrino mass matrix. 
Otherwise they give the off-diagonal elements.

As a simple example,
let us consider the one-loop level diagram shown in Fig.\ref{one-loop}. 
One can obtain similar diagram by replacing 
$e \to \nu,~\Delta^+ \to \Delta^0,~\Phi^+ \to \Phi^0$ in the loop.
These diagrams contribute only to off-diagonal elements of $M_{\nu}$, and its  
magnitude is at most about 
$(1/(16 \pi^2)) y_{\Delta}y_{\ell} m_{\tau}={\cal O}(10^{-4})~\eV$. 
This is enough smaller than the tree-level contribution $y_{\Delta}v_{\Delta}=
{\cal O}(0.1)~\eV$ because chirality suppression occurs.
As a result, one finds that it is negligible.

As another example, let us consider a contribution to diagonal elements from the
two-loop diagrams depicted in Fig.\ref{two-loop-1} and Fig.\ref{two-loop-2}, in which the dominant one comes from the upper-right panel in Fig.\ref{two-loop-1}.
However the magnitude is at most $(1/(16 \pi^2)^2)y_{\Delta}^3 \mu = {\cal O}(10^{-22})~\eV$ if 
$\mu ={\cal O}(1)~\TeV$, which is also tiny enough. 
Therefore we neglect contributions 
from those diagrams at all. Lastly, 
we have a three-loop contribution shown by Fig.\ref{three-loop} that gives off-diagonal elements, and we find that this is also too tiny to generate neutrino masses.  See Ref.\cite{Gustafsson} for details.

\end{appendix}


\clearpage

\begin{thebibliography}{99}

\bibitem{ATLAS_Higgs}
  G.~Aad {\it et al.}  [ATLAS Collaboration],
  Phys.\ Lett.\ B {\bf 716}, 1 (2012)
  [arXiv:1207.7214 [hep-ex]].


\bibitem{CMS_Higgs} 
  S.~Chatrchyan {\it et al.}  [CMS Collaboration],
  Phys.\ Lett.\ B {\bf 716}, 30 (2012)
  [arXiv:1207.7235 [hep-ex]].

\bibitem{Ishimori:2012zz} 
  H.~Ishimori, T.~Kobayashi, H.~Ohki, H.~Okada, Y.~Shimizu and M.~Tanimoto,
  Lect.\ Notes Phys.\  {\bf 858}, 1 (2012).

\bibitem{Ishimori:2010au} 
  H.~Ishimori, T.~Kobayashi, H.~Ohki, Y.~Shimizu, H.~Okada and M.~Tanimoto,
  Prog.\ Theor.\ Phys.\ Suppl.\  {\bf 183}, 1 (2010)
  [arXiv:1003.3552 [hep-th]].

\bibitem{Altarelli:2010gt} 
  G.~Altarelli and F.~Feruglio,
  Rev.\ Mod.\ Phys.\  {\bf 82}, 2701 (2010)
  [arXiv:1002.0211 [hep-ph]].


\bibitem{Babu:2002dz} 
  K.~S.~Babu, E.~Ma and J.~W.~F.~Valle,
  Phys.\ Lett.\ B {\bf 552}, 207 (2003)
  [hep-ph/0206292].

\bibitem{Ma:2001dn} 
  E.~Ma and G.~Rajasekaran,
  Phys.\ Rev.\ D {\bf 64}, 113012 (2001)
  [hep-ph/0106291].

\bibitem{Ferreira:2013oga} 
  P.~M.~Ferreira, L.~Lavoura and P.~O.~Ludl,
  arXiv:1306.1500 [hep-ph].

\bibitem{Luhn:2007sy} 
  C.~Luhn, S.~Nasri and P.~Ramond,
  Phys.\ Lett.\ B {\bf 652}, 27 (2007)
  [arXiv:0706.2341 [hep-ph]].

\bibitem{Ma:2007wu} 
  E.~Ma,
  Phys.\ Lett.\ B {\bf 660}, 505 (2008)
  [arXiv:0709.0507 [hep-ph]].

\bibitem{Ma:2006ip} 
  E.~Ma,
  Mod.\ Phys.\ Lett.\ A {\bf 21}, 1917 (2006)
  [hep-ph/0607056].

\bibitem{Ishimori:2012gv} 
  H.~Ishimori and T.~Kobayashi,
  Phys.\ Rev.\ D {\bf 85}, 125004 (2012)
  [arXiv:1201.3429 [hep-ph]].

\bibitem{Ma:2006ht} 
  E.~Ma,
  Phys.\ Lett.\ B {\bf 649}, 287 (2007)
  [hep-ph/0612022].





\bibitem{Ishimori:2012sw} 
  H.~Ishimori, S.~Khalil and E.~Ma,
  Phys.\ Rev.\ D {\bf 86}, 013008 (2012)
  [arXiv:1204.2705 [hep-ph]].

\bibitem{Cao:2011cp} 
  Q.~-H.~Cao, S.~Khalil, E.~Ma and H.~Okada,
  Phys.\ Rev.\ D {\bf 84}, 071302 (2011)
  [arXiv:1108.0570 [hep-ph]].

\bibitem{Cao:2010mp} 
  Q.~-H.~Cao, S.~Khalil, E.~Ma and H.~Okada,
  Phys.\ Rev.\ Lett.\  {\bf 106}, 131801 (2011)
  [arXiv:1009.5415 [hep-ph]].

\bibitem{Hagedorn:2008bc} 
  C.~Hagedorn, M.~A.~Schmidt and A.~Y.~.Smirnov,
  Phys.\ Rev.\ D {\bf 79}, 036002 (2009)
  [arXiv:0811.2955 [hep-ph]].

\bibitem{typeII} 
 T.~P.~Cheng and L.~F.~Li,
 Phys.\ Rev.\  D {\bf 22}, 2860 (1980);
 J.~Schechter and J.~W.~F.~Valle,
 Phys.\ Rev.\  D {\bf 22}, 2227 (1980);
  G.~Lazarides, Q.~Shafi and C.~Wetterich,
  Nucl.\ Phys.\  B {\bf 181}, 287 (1981);
  R.~N.~Mohapatra and G.~Senjanovic,
  Phys.\ Rev.\  D {\bf 23}, 165 (1981);
  M.~Magg and C.~Wetterich,
  Phys.\ Lett.\  B {\bf 94}, 61 (1980).


\bibitem{typeX} 
  M.~Aoki, S.~Kanemura, K.~Tsumura and K.~Yagyu,
  Phys.\ Rev.\ D {\bf 80}, 015017 (2009)
  [arXiv:0902.4665 [hep-ph]]. 

\bibitem{typeX2} 
  V.~Barger, H.~E.~Logan and G.~Shaughnessy,
  Phys.\ Rev.\ D {\bf 79}, 115018 (2009)
  [arXiv:0902.0170 [hep-ph]]; 
  H.~E.~Logan and D.~MacLennan,
  Phys.\ Rev.\ D {\bf 79}, 115022 (2009)
  [arXiv:0903.2246 [hep-ph]];
  H.~E.~Logan and D.~MacLennan,
  Phys.\ Rev.\ D {\bf 81}, 075016 (2010)
  [arXiv:1002.4916 [hep-ph]];
  G.~C.~Branco, P.~M.~Ferreira, L.~Lavoura, M.~N.~Rebelo, M.~Sher and J.~P.~Silva,
  Phys.\ Rept.\  {\bf 516}, 1 (2012)
  [arXiv:1106.0034 [hep-ph]].

 \bibitem{dimuon}
ATLAS-CONF-2013-010. 

  \bibitem{zee-babu}
A. Zee, {\em Nucl. Phys. B} {\bf 264}, 99 (1986);
K. S. Babu, {\em Phys. Lett. B} {\bf 203}, 132 (1988).

\bibitem{Ma:2006km}
  E.~Ma,
  Phys.\ Rev.\  D {\bf 73}, 077301 (2006)
  [arXiv:hep-ph/0601225].

\bibitem{Sahu}
 N.~Sahu and U.~Sarkar,
  Phys.\ Rev.\ D {\bf 78}, 115013 (2008)
  [arXiv:0804.2072 [hep-ph]].

\bibitem{Aoki:2013gzs} 
  M.~Aoki, J.~Kubo and H.~Takano,
  Phys.\  Rev.\  D 87, {\bf 116001} (2013)
  [arXiv:1302.3936 [hep-ph]].


\bibitem{Krauss:2002px}
  L.~M.~Krauss, S.~Nasri and M.~Trodden,
  Phys.\ Rev.\  D {\bf 67}, 085002 (2003)
  [arXiv:hep-ph/0210389].



\bibitem{Aoki:2008av}
  M.~Aoki, S.~Kanemura and O.~Seto,
  Phys.\ Rev.\ Lett.\  {\bf 102}, 051805 (2009)
  [arXiv:0807.0361]; 
  M.~Aoki, S.~Kanemura and O.~Seto,
  Phys.\ Rev.\ D {\bf 80}, 033007 (2009)
  [arXiv:0904.3829 [hep-ph]];
  M.~Aoki, S.~Kanemura and K.~Yagyu,
  Phys.\ Rev.\ D {\bf 83}, 075016 (2011)
  [arXiv:1102.3412 [hep-ph]].


\bibitem{Schmidt:2012yg} 
  D.~Schmidt, T.~Schwetz and T.~Toma,
  Phys.\ Rev.\ D {\bf 85}, 073009 (2012)
  [arXiv:1201.0906 [hep-ph]].

\bibitem{Bouchand:2012dx} 
  R.~Bouchand and A.~Merle,
  arXiv:1205.0008 [hep-ph].

\bibitem{Ma:2012ez} 
  E.~Ma, A.~Natale and A.~Rashed,
  arXiv:1206.1570 [hep-ph].




\bibitem{Kajiyama:2013zla} 
  Y.~Kajiyama, H.~Okada and K.~Yagyu,
    Nucl.\ Phys.\ B {\bf 874}, 1 (2013)
  arXiv:1303.3463 [hep-ph].

\bibitem{Aoki:2011he} 
  M.~Aoki, J.~Kubo, T.~Okawa and H.~Takano,
  Phys.\ Lett.\ B {\bf 707}, 107 (2012)
  [arXiv:1110.5403 [hep-ph]].

\bibitem{Ahn:2012cg} 
  Y.~H.~Ahn and H.~Okada,
  Phys.\ Rev.\ D {\bf 85}, 073010 (2012)
  [arXiv:1201.4436 [hep-ph]].

\bibitem{Farzan:2012sa} 
  Y.~Farzan and E.~Ma,
  arXiv:1204.4890 [hep-ph].

\bibitem{Bonnet:2012kz} 
  F.~Bonnet, M.~Hirsch, T.~Ota and W.~Winter,
  arXiv:1204.5862 [hep-ph].

\bibitem{Kumericki:2012bf} 
  K.~Kumericki, I.~Picek and B.~Radovcic,
  arXiv:1204.6597 [hep-ph].

\bibitem{Kumericki:2012bh} 
  K.~Kumericki, I.~Picek and B.~Radovcic,
  arXiv:1204.6599 [hep-ph].

\bibitem{Ma:2012if} 
  E.~Ma,
  arXiv:1206.1812 [hep-ph].


\bibitem{Gil:2012ya} 
  G.~Gil, P.~Chankowski and M.~Krawczyk,
  arXiv:1207.0084 [hep-ph].

\bibitem{Okada:2012np} 
  H.~Okada and T.~Toma,
  Phys.\ Rev.\ D {\bf 86}, 033011 (2012)
  arXiv:1207.0864 [hep-ph].

\bibitem{Hehn:2012kz} 
  D.~Hehn and A.~Ibarra,
  Phys.\ Lett.\ B {\bf 718}, 988 (2013)
  [arXiv:1208.3162 [hep-ph]].

\bibitem{Dev:2012sg} 
  P.~S.~B.~Dev and A.~Pilaftsis,
  Phys.\ Rev.\ D {\bf 86}, 113001 (2012)
  [arXiv:1209.4051 [hep-ph]].

\bibitem{Kajiyama:2012xg} 
  Y.~Kajiyama, H.~Okada and T.~Toma,
  arXiv:1210.2305 [hep-ph].

\bibitem{Okada:2012sp} 
  H.~Okada,
  arXiv:1212.0492 [hep-ph].



\bibitem{Aoki:2010ib} 
  M.~Aoki, S.~Kanemura, T.~Shindou and K.~Yagyu,
  JHEP {\bf 1007}, 084 (2010)
  [Erratum-ibid.\  {\bf 1011}, 049 (2010)]
  [arXiv:1005.5159 [hep-ph]].

\bibitem{Kanemura:2011vm} 
  S.~Kanemura, O.~Seto and T.~Shimomura,
  Phys.\ Rev.\ D {\bf 84}, 016004 (2011)
  [arXiv:1101.5713 [hep-ph]].

\bibitem{Lindner:2011it} 
  M.~Lindner, D.~Schmidt and T.~Schwetz,
  Phys.\ Lett.\ B {\bf 705}, 324 (2011)
  [arXiv:1105.4626 [hep-ph]].

\bibitem{Kanemura:2011mw1} 
  S.~Kanemura, T.~Nabeshima and H.~Sugiyama,
  Phys.\ Rev.\ D {\bf 85}, 033004 (2012)
  [arXiv:1111.0599 [hep-ph]].

\bibitem{Kanemura:2011mw2} 
  S.~Kanemura, T.~Nabeshima and H.~Sugiyama,
  Phys.\ Rev.\ D {\bf 85}, 033004 (2012)
  [arXiv:1111.0599 [hep-ph]].

\bibitem{Kanemura_Sugiyama} 
  S.~Kanemura and H.~Sugiyama,
  Phys.\ Rev.\ D {\bf 86}, 073006 (2012)
  [arXiv:1202.5231 [hep-ph]].


\bibitem{Gu:2007ug} 
  P.~-H.~Gu and U.~Sarkar,
  Phys.\ Rev.\ D {\bf 77}, 105031 (2008)
  [arXiv:0712.2933 [hep-ph]].

\bibitem{Gu:2008zf} 
  P.~-H.~Gu and U.~Sarkar,
  Phys.\ Rev.\ D {\bf 78}, 073012 (2008)
  [arXiv:0807.0270 [hep-ph]].

\bibitem{Gustafsson} 
M.~Gustafsson, J.~M.~No and M.~A.~Rivera,
  Phys.\ Rev.\ Lett.\  {\bf 110}, 211802 (2013)
arXiv:1212.4806 [hep-ph]. 


\bibitem{two-triplet}  
  Y.~Kajiyama, H.~Okada and T.~Toma,
  arXiv:1303.7356 [hep-ph].

\bibitem{Kanemura:2013qva} 
  S.~Kanemura, T.~Matsui and H.~Sugiyama,
  arXiv:1305.4521 [hep-ph].

\bibitem{Law:2013saa} 
  S.~S.~C.~Law and K.~L.~McDonald,
  arXiv:1305.6467 [hep-ph].


\bibitem{PDG}
Beringer et al. (Particle Data Group),~Phys.\ Rev.\ D~{\bf 86}, 010001 (2012).

\bibitem{Misiak} 
  M.~Misiak, H.~M.~Asatrian, K.~Bieri, M.~Czakon, A.~Czarnecki, T.~Ewerth, A.~Ferroglia and P.~Gambino {\it et al.},
  Phys.\ Rev.\ Lett.\  {\bf 98}, 022002 (2007)
  [hep-ph/0609232].

\bibitem{Stal} 
  F.~Mahmoudi and O.~Stal,
  Phys.\ Rev.\ D {\bf 81}, 035016 (2010)
  [arXiv:0907.1791 [hep-ph]].

\bibitem{Babar} 
  J.~P.~Lees {\it et al.}  [BaBar Collaboration],
  Phys.\ Rev.\ Lett.\  {\bf 109}, 101802 (2012)
  [arXiv:1205.5442 [hep-ex]].


\bibitem{Moriond_ATLAS}
ATLAS-CONF-2013-012;
ATLAS-CONF-2013-013;
ATLAS-CONF-2013-030;
ATLAS-CONF-2012-170.

\bibitem{Moriond_CMS} 
CMS-PAS-HIG-13-005;
CMS-PAS-HIG-13-001;
CMS-PAS-HIG-13-002;
CMS-PAS-HIG-13-003;
CMS-PAS-HIG-13-004.


\bibitem{MSSM_neutral_ATLAS}
  G.~Aad {\it et al.}  [ATLAS Collaboration],
  JHEP {\bf 1302}, 095 (2013)
  [arXiv:1211.6956 [hep-ex]].


\bibitem{MSSM_neutral_CMS}
  G.~Aad {\it et al.}  [ATLAS Collaboration],
  JHEP {\bf 1302}, 095 (2013)
  [arXiv:1211.6956 [hep-ex]].


\bibitem{MSSM_charged_ATLAS} 
  G.~Aad {\it et al.}  [ATLAS Collaboration],
  JHEP {\bf 1206}, 039 (2012).

\bibitem{MSSM_charged_CMS} 
  S.~Chatrchyan {\it et al.}  [CMS Collaboration],
  JHEP {\bf 1207}, 143 (2012). 

 







\bibitem{Higgs_cross} 
https://twiki.cern.ch/twiki/bin/view/LHCPhysics/CERNYellowReportPageAt7TeV

\bibitem{MG5} 
  J.~Alwall, M.~Herquet, F.~Maltoni, O.~Mattelaer and T.~Stelzer,
  JHEP {\bf 1106}, 128 (2011).

\bibitem{Yokoya} 
  S.~Kanemura, K.~Tsumura and H.~Yokoya,
  Phys.\ Rev.\ D {\bf 85}, 095001 (2012)
  [arXiv:1111.6089 [hep-ph]];
  S.~Kanemura, K.~Tsumura and H.~Yokoya,
  arXiv:1305.5424 [hep-ph].

\bibitem{CalcHEP}
 A.~Pukhov,
  [hep-ph/0412191].  


\end{thebibliography}
\end{document}